\documentclass[twocolumn]{aastex63}

\shorttitle{The BoRG-JWST UV Luminosity Function at $z>7$}
\shortauthors{Rojas-Ruiz et al.}

\usepackage{float}
\usepackage{natbib}
\usepackage{hyperref}
\usepackage[nointegrals]{wasysym}
\usepackage{ragged2e}
\usepackage{graphicx}
\usepackage{units}
\usepackage{amsmath}
\definecolor{red}{rgb}{1,0,0}
\definecolor{orange}{RGB}{204, 85, 0}
\definecolor{blue}{HTML}{4169e1}
\definecolor{ltred}{RGB}{245,167,162}
\definecolor{ltblue}{RGB}{206,211,242}

\newcommand{\hst}{\textit{HST}}
\newcommand{\jwst}{\textit{JWST}}

\defcitealias{roberts-borsani_borg-jwst_2024}{RB24}
\defcitealias{rojas-ruiz_probing_2020}{RR20}
\defcitealias{leethochawalit_uv_2023}{L23}
\begin{document}

\title{The BoRG-\textit{JWST} Survey: Abundance and Mass-to-light Ratio of Luminous $z=7-9$ Galaxies from Independent Sight Lines with NIRSpec}

\correspondingauthor{Sof\'ia Rojas-Ruiz}
\email{rojas@astro.ucla.edu}

\author[0000-0003-2349-9310]{Sof\'ia Rojas-Ruiz}\affiliation{Department of Physics and Astronomy, University of California, Los Angeles, 430 Portola Plaza, Los Angeles, CA 90095, USA}

\author[0000-0002-9921-9218]{Micaela Bagley}
\affiliation{Department of Astronomy, The University of Texas at Austin, 2515 Speedway, Austin, TX, 78712, USA}\textbf{}

\author[0000-0002-4140-1367]{Guido Roberts-Borsani}
 \affiliation{Department of Astronomy, University of Geneva, Chemin Pegasi 51, 1290 Versoix, Switzerland}

\author[0000-0002-8460-0390]{Tommaso Treu}
\affiliation{Department of Physics and Astronomy, University of California, Los Angeles, 430 Portola Plaza, Los Angeles, CA 90095, USA}

\author[0000-0001-8519-1130]{Steven L. Finkelstein}
\affiliation{Department of Astronomy, The University of Texas at Austin, 2515 Speedway, Austin, TX, 78712, USA}

\author[0000-0002-8512-1404]{Takahiro Morishita}
\affiliation{IPAC, California Institute of Technology, MC 314-6, 1200 E. California Boulevard, Pasadena, CA 91125, USA}

\author[0000-0003-4570-3159]{Nicha Leethochawalit}
\affiliation{National Astronomical Research Institute of Thailand (NARIT), Mae Rim, Chiang Mai, 50180, Thailand}

\author[0000-0002-3407-1785]{Charlotte Mason}
\affiliation{Cosmic Dawn Center (DAWN)}
\affiliation{Niels Bohr Institute, University of Copenhagen, Jagtvej 128, DK-2200, Copenhagen N, Denmark}

\author[0000-0002-2931-7824]{Eduardo Ba{\~n}ados}
\affiliation{Max Planck Institut f\"ur Astronomie, K\"onigstuhl 17, D-69117, Heidelberg, Germany}

\author[0000-0001-9391-305X]{Michele Trenti}
\affiliation{School of Physics, University of Melbourne, Parkville 3010, VIC, Australia}
\affiliation{ARC Centre of Excellence for All Sky Astrophysics in 3 Dimensions (ASTRO 3D), Australia}

\author[0000-0001-9935-6047]{Massimo Stiavelli}
\affiliation{Space Telescope Science Institute, 3700 San Martin Drive, Baltimore, MD 21218, USA}
\affiliation{Dept. of Physics \& Astronomy, Johns Hopkins University, Baltimore, MD 21218, USA}
\affiliation{Dept. of Astronomy, University of Maryland, College Park, MD 20742, USA}

\author[0000-0003-3466-035X]{{L. Y. Aaron} {Yung}}
\affiliation{Space Telescope Science Institute, 3700 San Martin Dr., Baltimore, MD 21218, USA}

\author[0000-0002-7959-8783]{Pablo Arrabal Haro}
\affiliation{NSF’s National Optical-Infrared Astronomy Research Laboratory, 950 N. Cherry Ave., Tucson, AZ 85719, USA}

\author[0000-0002-6748-6821]{Rachel S. Somerville}
\affiliation{Center for Computational Astrophysics, Flatiron Institute, 162 5th Avenue, New York, NY 10010, USA}

\author[0000-0001-7840-2972]{Christian Soto}
\affiliation{Space Telescope Science Institute, 3700 San Martin Drive, Baltimore, MD 21218, USA}

%=====================================================
%============ Abstract ===============================
%====================================================
\begin{abstract}
We present new results on the rest-frame UV luminosity function (UVLF) and stellar mass-to-light (M/L) ratio of bright (M$_{\rm UV}\lesssim-20$ mag) spectroscopically-confirmed galaxies at $z=7-9$ derived from the BoRG-\jwst\ survey, a unique data set of NIRSpec prism follow-up of \textit{Hubble Space Telescope (HST)--}selected sources from random-pointing imaging.
By selecting galaxies from over 300 independent sight lines, the survey minimizes cosmic variance ensuring a statistically robust sample of the bright-galaxy population during the epoch of reionization. The data are used to constrain, for the first time, the bright end of the UVLF at $z=7-9$ from spectroscopically confirmed galaxies over eight independent fields. We find that the bright end of the UVLF is higher than found using imaging over \jwst\ legacy fields, suggesting that the latter may be significantly affected by cosmic variance, and thus reducing the tension with recent findings from \jwst\ at $z>10$ and comparable to models invoking little dust attenuation and bursty star formation. Additionally, we use the galaxies' \jwst\ spectra to infer their stellar masses and M/L ratios relative to other \hst\ and \jwst\ studies. We show that the stellar mass scales almost linearly with UV luminosity (M$_* \propto L_{\rm UV}^{0.85\pm0.12}$), albeit with large ($\sim0.5$ dex) intrinsic scatter, consistent with stochastic bursts of star formation in early galaxy formation.

\end{abstract}

\keywords{galaxies: high-redshift, star formation -  cosmology: observations, reionization, dark ages}
\vspace{0.5cm}

\section{Introduction}\label{intro}
The abundance of UV-bright galaxies is a key quantity for understanding the critical physical processes shaping galaxy evolution, particularly at early cosmic times, from cosmic dawn to the epoch of hydrogen reionization \citep[see][and references within]{robertson_galaxy_2022}. The rest-frame UV luminosity function (UVLF), which details the number density of UV-bright galaxies, is a powerful probe of the intricate dynamics responsible for star formation efficiency driven by supernova and active galactic nucleus (AGN) feedback, dust attenuation, and the buildup of dark matter halos (e.g. \citealt{somerville_semi-analytic_2008, bower_what_2012, finkelstein_candels_2012}).

The \textit{Hubble Space Telescope (HST)} has been used for over a decade to find UV-bright galaxy candidates when the Universe was as young as $\sim$ 400 million years old ($z\sim11$). Deep legacy surveys pioneering this search include the Cosmic Assembly Near-infrared Deep Extragalactic Legacy Survey (CANDELS; \citealt{grogin_candels_2011,koekemoer_candels_2011}, the Hubble Ultra Deep Field \citealt{bouwens_discovery_2010,oesch_z_2010,ellis_abundance_2013}, and the Hubble Frontier Fields (HFF; \citealt{lotz_frontier_2017}). 

A complimentary method to find bright galaxy candidates that may not be fully sampled by deep surveys covering the same area in contiguous fields are the pure-parallel programs such as the WFC3 Infrared Spectroscopic Parallel survey \citep[WISP;][]{atek_wfc3_2010}, the Brightest of Reionizing Galaxies Survey \citep[BoRG;][]{trenti_brightest_2011}, and the Hubble Infrared Pure Parallel Imaging Extragalactic Survey \citep[HIPPIES;][]{yan_probing_2011}. The pure-parallel observations provide completely independent, uncorrelated observations that reduce the uncertainty due to cosmic variance from $\gtrsim$20\% to $<$1\%, thus providing an ideal complement to similar-depth surveys in a single pointing 
(\citealt{trenti_cosmic_2008,bradley_brightest_2012,trenti_overdensities_2012,schmidt_luminosity_2014,trapp_framework_2022}).

Despite the wealth of data available, the density of UV-bright galaxies at early cosmic times is ill constrained by \hst\ alone. The limitation arises because at $z >8$ only a few \hst\ filters (F105W, F140W, and F160W hereafter) sample the wavelength range redward of the $\lambda_{rest} = 1216$ \AA\ Ly$\alpha$ line. Some of the candidates are even detected in only one filter \citep[e.g.,][]{coe_clash_2013,oesch_most_2014,finkelstein_census_2022}. The limited wavelength range available with HST makes some degree of contamination by lower-redshift interlopers unavoidable. Frequent interlopers are brown dwarfs, which share a similar $\sim1-2\ \mu$m color-color space given their low temperatures, and $z\sim1-3$ galaxies with a strong rest-frame $4000$ \AA\ Balmer break caused by the accumulation of absorption lines from ionized metals and that can be confused for the Lyman break of $z=7-10$ galaxies. To accurately constrain the rest-UV luminosity function at $z>7$, it is crucial to confirm the high-redshift nature of the galaxies, which was extremely difficult prior to \jwst\ \citep[e.g.,][]{livermore_hst_2018,mason_inferences_2019,larson_searching_2022}.

Within this context, studies of the UVLF from pure-parallel programs hinted at an excess of bright ($M_{UV}<-21$) galaxies with respect to those found in legacy surveys or extrapolated by fitting a Schechter functions or a double power law to the more abundant fainter galaxies \citep[][]{bernard_galaxy_2016,calvi_bright_2016,morishita_bright-end_2018,rojas-ruiz_probing_2020,leethochawalit_uv_2023,bagley_bright_2024}. These results favored a smooth evolution of the UVLF at the bright end \citep[e.g.][]{mcleod_new_2015,finkelstein_observational_2016,finkelstein_census_2022,mcleod_z_2016} compared to a rapid decline suggested by other studies \citep[e.g.][]{oesch_probing_2013, oesch_most_2014,bouwens_uv_2015,bouwens_bright_2016,oesch_dearth_2018,bouwens_newly_2019, bouwens_new_2021,bowler_lack_2020}. However, prior to \jwst\ it was not possible to determine whether this ``bright excess" was real or the result of larger contamination from a lower-redshift interloper than estimated. As a result, consensus on the rest-UV luminosity functions at $z>8$ and the physical processes in early galaxy formation remained elusive. 

The unparalleled sensitivity and resolution of \jwst\ to find high-redshift galaxies with the Near-Infrared Camera NIRCam (0.6 -- 5 $\mu$m) are revolutionizing the study of the rest-UV luminosity function. While \jwst\ is not optimized to conduct wide-field surveys, it has targeted previous \hst\ legacy fields with NIRCam to look for new UV-bright, high-redshift galaxies. Certain fields provide deep imaging to improve sample purity such as the GLASS-JWST Early Release Science \citep{treu_glass-jwst_2022}, the \jwst\ Advanced Deep Extragalactic Survey \citep[JADES;][]{eisenstein_overview_2023}, the Cosmic Evolution Early Release Science field \citep[CEERS;][]{finkelstein_complete_2024},  the First Reionization Epoch Spectroscopically Complete Observations \citep{oesch_jwst_2023}, the Prime Extragalactic Areas for Reionization and Lensing Science \citep[PEARLS;][]{windhorst_jwst_2023}, the Next Generation Deep Extragalactic Exploratory Public survey \citep[NGDEEP;][]{bagley_next_2024}, and the Public Release IMaging for Extragalactic Research (PRIMER; PI: Dunlop). Other programs that offer a larger search field to reduce the effects of cosmic variance, although with the caveat of surveying shallower imaging and risking a higher contamination fraction, are the COSMOS-Web \citep{casey_cosmos-web_2023} program, the Pure Parallel Wide Area Legacy Imaging with JWST/NIRCam  \citep[PANORAMIC;][]{williams_panoramic_2025}, and the Bias-free Extragalactic Analysis for Cosmic Origins with NIRCam \citep[BEACON;][]{morishita_beacon_2025}.

These \jwst\ surveys not only have improved the purity of galaxy  catalogs at $z=7-10$ by virtue of the longer-wavelength coverage, but also have explored for the first time $z\gtrsim10$ galaxies to study the evolution of the UVLF \citep[e.g.,][]{bouwens_uv_2023,bouwens_evolution_2023,donnan_evolution_2023,donnan_jwst_2024,finkelstein_complete_2024,leung_ngdeep_2023,adams_epochs_2024,casey_cosmos-web_2024,harikane_pure_2024,harikane_jwst_2025}. However, the great majority of these \jwst\ UVLFs are constructed from catalogs of photometrically selected galaxies and can be contaminated by lower-redshift sources. The true power of \jwst\ lies in targeted follow-up for redshift confirmation and source characterization of UV-luminous galaxies (e.g., \citealt{arrabal_haro_confirmation_2023,carniani_spectroscopic_2024,castellano_jwst_2024,roberts-borsani_between_2024}). The Near-Infrared Spectrograph NIRSpec (0.6-5 $\mu$m) can effectively distinguish the $z\gtrsim7$ galaxies from lower-redshift interlopers. 

In this study, we leverage NIRSpec observations from the BoRG-\jwst\ survey \citep{roberts-borsani_borg-jwst_2025} that confirm 12 galaxies at $z=7-10$ that were previously selected from the \hst\ surveys SuperBoRG, BoRG, HIPPIES, and WISP \citep{rojas-ruiz_probing_2020, morishita_superborg_2021, bagley_bright_2024}. This is the largest pure-parallel sample yet analyzed that comprises some of the brightest galaxies at these redshifts. As we will show, our studies confirm that the bright excess found by \hst\ pure-parallel programs is confirmed and not due to higher contamination than expected. The confirmation of the excess reduces the amount of evolution required to match the abundance of bright galaxies found by \jwst\ at higher redshift. Furthermore, we find that the stellar mass correlates with UV luminosity, albeit with a large scatter, consistent with stochastic early star formation. 

This paper is organized as follows: In \S \ref{sec-lf}, we present the bright end of the rest-frame ultraviolet luminosity functions (UVLFs) from our BoRG-\jwst\ survey, and compare them to other studies from observations in \S \ref{comp_obs}. We further compare our results to theoretical models with different recipes of physical processes in galaxies that build up the UVLF \S \ref{sims-lf}. We discuss the process for estimating the stellar masses of the galaxies and the mass-to-light ratio relation in \S \ref{sec-stellar}. Finally, the summary and implications of our findings are discussed in \S \ref{discuss}. Throughout this work we use the cosmology according to H$_0$= 70 km~s$^{-1}$ Mpc$^{-1}$, $\Omega_M$= 0.3, and $\Omega_\Lambda$ = 0.7 and a \citet{chabrier_galactic_2003} initial mass function (IMF). All magnitudes are given in the AB system.

\begin{deluxetable*}{lccccc}
\tablecaption{UV luminosity functions for BoRG-JWST}
\tablewidth{700pt}
\tablehead{
\colhead{M$_{UV}$ }
& \colhead{No. Photometry} & \colhead{No. Observed} & \colhead{No. Confirmed}& \colhead{$\phi$} & %\colhead{$\partial \phi$} & 
\colhead{V$_{eff}$} \\
\colhead{} & \colhead{} & \colhead{} & \colhead{} &
\colhead{(10$^{-6}$ Mpc$^{-3}$ mag$^{-1}$)} &
\colhead{10$^4$ Mpc$^3$}
}
\startdata
\multicolumn{6}{c}{GO 1747 \textbf{($7.5 < z < 8.5$)}} \\
\hline
$-23.0$ & 1 & 1 & 0 & $<2.4096^*$ & 152.810\\
$-22.5$ & 0 & 0 & 0 & 1.6672$\pm$0.929 & 153.185\\
$-22.0$ & 7 & 2 & 0 & 5.2094$\pm$1.717 & 143.166\\
$-21.5$ & 11 & 4 & 4 & 24.773$\pm$4.372 & 75.659\\
$-21.0$ & 6 & 3 & 1 & 16.535$\pm$4.795 & 12.732\\
$-20.5$ & 1 & 0 & 1 & 244.90$\pm$139.2 & 0.817\\
\hline
\hline
\multicolumn{6}{c}{GO 2426 \textbf{($7.0 < z \leq8.4$)}} \\
\hline
$-23.0$ & 1 & 0 & 0 & 5.6899$^{+5.482}_{-3.270}$ & 29.364\\
$-22.0$ & 1 & 0 & 1 & 4.6464$^{+5.231}_{-2.918}$ & 27.604\\
$-21.0$ & 3 & 1 & 3 & 107.31$^{+54.76}_{-40.77}$ & 4.654\\
$-20.0$ & 3 & 3 & 0 & 399.37$^{+273.0}_{-271.6}$ & 0.264\\
\hline
\multicolumn{6}{c}{GO 2426 \textbf{($8.4 < z < 10$)}} \\
\hline
$-23.0$ & 0 & 0 & 0 & 0.9620$^{+9.347}_{-0.862}$ & 65.471\\
$-22.0$ & 1 & 1 & 0 & 2.8770$^{+0.711}_{-0.882}$ & 45.811\\
$-21.0$ & 1 & 1 & 1 & 20.805$^{+40.95}_{-11.10}$ & 9.127\\
$-20.0$ & 0 & 0 & 1 & 239.48$^{+133.2}_{-102.9}$ & 0.417\\
\enddata
\tablecomments{These number density values are  plotted in Figure \ref{uv-lf}. Column 1 presents the $M_{UV}$ bins, individual values for each galaxy are reported in Table \ref{mass_table}. Column 2 has the number of galaxies from the photometric samples in each $M_{UV}$ bin from \citetalias{leethochawalit_uv_2023} and \citetalias{rojas-ruiz_probing_2020}. Column 3 is the number of galaxies from the photometric sample observed in the GO 1747 and GO 2426 programs, respectively. Column 4 shows the number of galaxies confirmed with NIRSpec. Note that a few galaxies changed the redshift and $M_{UV}$ bins when confirmed. Column 5 shows the calculated number density with 68\% interval uncertainties using the photometric sample corrected by the estimated rescaled contamination given the spectroscopically confirmed galaxies. Column 6 show the effective volume as calculated in \citetalias{leethochawalit_uv_2023} and \citetalias{rojas-ruiz_probing_2020}, respectively.}
\tablenotetext{*}{$1\sigma$ errors from \citealt{gehrels_confidence_1986}.}
\label{lf_table}
\end{deluxetable*}

\section{The \lowercase{$z\sim8$} and \lowercase{$z\sim9$} UVLF\lowercase{s} from Independent Sight lines}\label{sec-lf}

We use data from the BoRG-\jwst\ survey -- the combination of \jwst\ Cycle 1 programs GO 1747 (PI: Roberts-Borsani) and GO 2426 (Co-PIs: Bagley and Rojas-Ruiz) -- to recalculate the $z\sim8$ and $z\sim9$ galaxy UVLFs previously derived using \hst\ imaging alone by \cite{rojas-ruiz_probing_2020} and \citet{leethochawalit_uv_2023} (hereafter \citetalias{rojas-ruiz_probing_2020} and \citetalias{leethochawalit_uv_2023}, respectively). 
Specifically, we employ the redshifts and UV absolute magnitudes of the 12 confirmed $z=7-9$ sources from Table 1 of \citet[][hereafter \citetalias{roberts-borsani_borg-jwst_2025}]{roberts-borsani_borg-jwst_2025} resulting from NIRSpec observations over nine independent fields, and the photometric candidates that were not targeted but are part of the UVLFs from \citetalias{rojas-ruiz_probing_2020} and \citetalias{leethochawalit_uv_2023}. For consistency, we adopt the redshift bin division of \citetalias{rojas-ruiz_probing_2020}, namely $7.0<z\leq8.4$ and $8.4<z<10$ for the GO 2426 $z\sim8$ and $z\sim9$ UVLFs, respectively, and $7.5<z<8.5$ from \citetalias{leethochawalit_uv_2023} for the GO 1747 $z\sim8$ UVLF. Below we describe the UVLF calculations for the corresponding datasets of the BoRG-\jwst\ survey. For reference, in addition to providing our best estimates of the UVLF, we also present the hard lower limits on the number densities obtained by counting only spectroscopically confirmed galaxies, divided by the full survey volume. We note that these are very conservative because our spectroscopic follow-up was restricted to a subset of less than half of the photometric candidates \citep{roberts-borsani_borg-jwst_2025}.

\subsection{UVLF Calculation for GO 1747}\label{go1747}
The $z\sim8$ UVLF presented in \citetalias{leethochawalit_uv_2023} was derived based on the \hst\ sample of color- and photo-$z$-selected sources from \citet{roberts-borsani_physical_2022}, the latter of which served as the parent sample for the GO 1747 program. Six of the primary $z>7$ targets from that program were confirmed (see Table 1 of \citetalias{roberts-borsani_borg-jwst_2025}). To recalculate the UVLF at $z\sim8$ from this program, we follow the methodology described by \citetalias{leethochawalit_uv_2023} where the estimated contamination probability of a true high-z galaxy is given by the following equation:

\begin{equation}
     P_{\rm high-z} = P_{\rm gal}\times(1-\overline{N}_{\rm interlopers})
\end{equation}

where $P_{\rm gal}$ is the probability that the source is not a brown dwarf, and $\overline{N}_{\rm interlopers}$ is the estimated number of interlopers (lower-redshift galaxy) in a BoRG field (see Table 2 in \citetalias{leethochawalit_uv_2023}). In our approach to update the UVLF, we replace $P_{\rm high-z}$ of the sources observed with NIRSpec with the actual measured values (0 for confirmed interlopers, and 1 for confirmed high-z galaxies). 

We also use the spectroscopic sample to evaluate the performance of our estimated contamination probability based on the photometry. The average probability of the contamination from the NIRSpec targets ($\overline{P}_{\rm spec}$) is 0.55, while the average probability from photometric candidates in \citetalias{leethochawalit_uv_2023} ($\overline{P}_{L23}$) is 0.74$^{+0.14}_{-0.23}$. These values are consistent within 1$\sigma$ with the spectroscopic measurement. To be conservative, for the galaxies that were not targeted for spectroscopic follow-up, we adjust the contamination probability estimates by rescaling them by the factor needed to match exactly the average contamination of the spectroscopic sample:
\begin{equation}
     P_{\rm high-z} = P_{\rm gal}\times\frac{\overline{P}_{\rm spec}}{\overline{P}_{L23}}
\end{equation}
The correction factor is small (0.74) and does not affect our conclusions in any way. 

To derive the updated UVLF, we perform 100 Monte Carlo iterations based on the \citetalias{leethochawalit_uv_2023} method. This involves drawing galaxies first according to the contamination probability ($P_{\rm high-z}$) and then from the interloper probability ($P = e^{-\overline{N}_{\rm interloper}}$). The resulting number densities and volumes are presented in Table~\ref{lf_table}, and in the left panel of Figure \ref{uv-lf} they are shown as filled orange diamonds, while the \citetalias{leethochawalit_uv_2023} number densities are shown as black squares.

\subsection{UVLF Calculation for GO 2426}\label{go2426}
The GO 2426 program selected $z \sim 8-10$ galaxies for spectroscopic follow up on the basis of a photo-$z$ analysis of pure-parallel \hst\ WFC3 imaging using the parent sample from \citetalias{rojas-ruiz_probing_2020} and the similarly-derived sample from \citet{bagley_bright_2024}. Two sources observed in this program were inconclusive (see \citetalias{roberts-borsani_borg-jwst_2025}) and pertained to the sample of \citet{bagley_bright_2024}. Since we do not use these sources, we simply update the UVLFs of \citetalias{rojas-ruiz_probing_2020} now including four confirmed galaxies at $z\sim8$, and two confirmations at $z\sim9$. 

\begin{figure*}
\centering
\includegraphics[width=\linewidth]{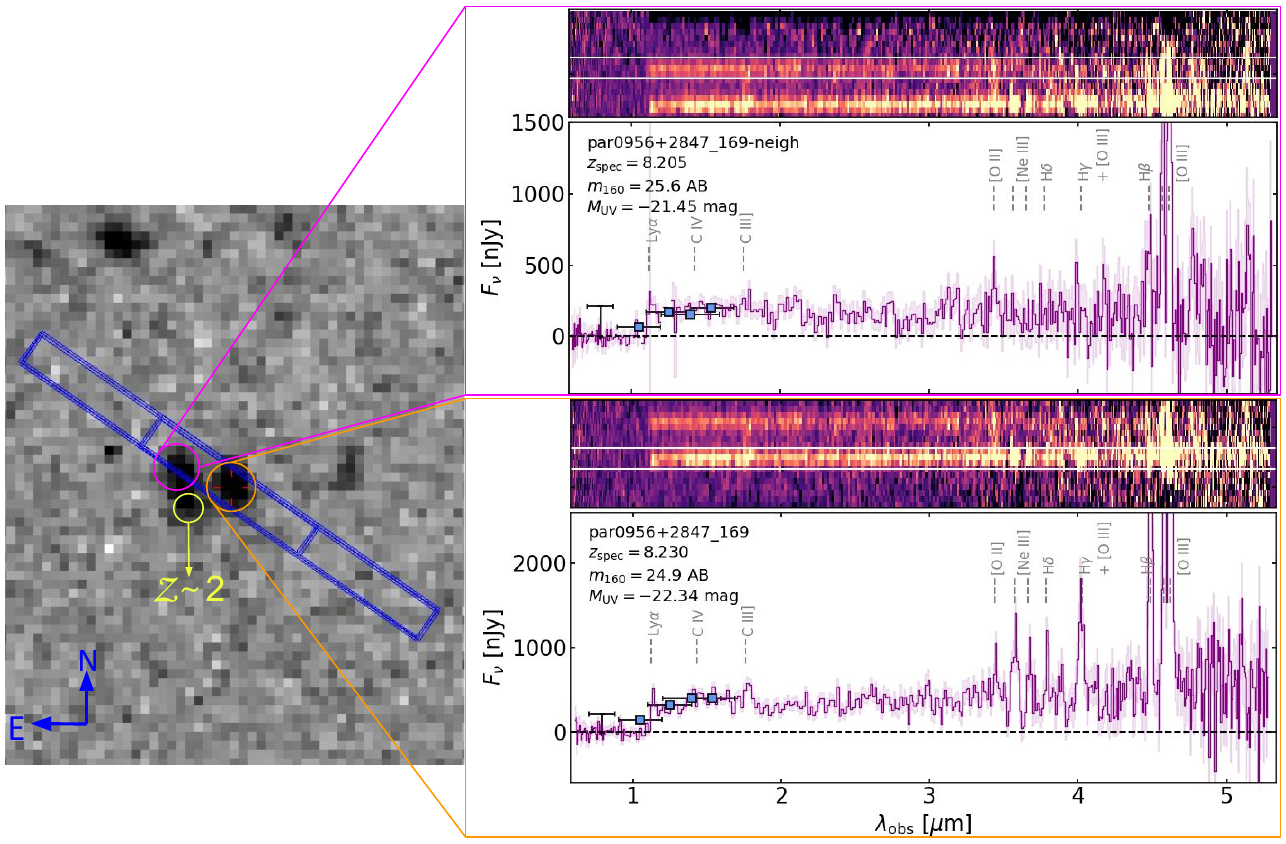}
\caption{\textit{Left: }The \hst\ image in the F160W filter of the primary target galaxy par0956$+$2847\_169 ($z=8.230$, orange) and the newly identified neighboring galaxy par0956$+$2847\_169-neigh ($z=8.205$, magenta). The \jwst-NIRSpec slit and dithers are shown in blue at a position angle PA = 55.07$^\circ$. Both galaxies are located at close separation of $\sim0\farcs63$ equivalent to $\sim$2.9 kpc and a relative line-of-sight velocity of $v_r=812$ km~s$^{-1}$ suggesting a merging system. \textit{Right:} The NIRSpec spectra of both galaxies, highlighting all the rest-frame optical emission lines identified in gray, including strong Ly-$\alpha$. The \hst\ photometry is also included as blue squares.}
\label{fig169}
\end{figure*}

It is worth noting that for the $z\sim8$ UVLF we include par0956$+$2847\_169-neigh, a newly identified bright galaxy ($z=8.205$, M$_{UV} = -21.45\pm0.09$) in close separation ($\sim0\farcs63$, or $\sim$2.9 kpc) of the main target and confirmed galaxy par0956$+$2847\_169 ($z=8.230$, M$_{UV}=-22.34\pm0.06$). Figure \ref{fig169} shows their spectra and relative positions. The relative lin-of-sight velocity $v_r = 812$ km~s$^{-1}$ suggests that these galaxies are gravitationally bound and likely a merger. We caution the following regarding interpretation of the bright end of the luminosity function (see examples by \citealt{bowler_unveiling_2017}, \citealt{dalmasso_rate_2024}, \citealt{duan_galaxy_2025}, and \citealt{adamo_first_2024}): counting mergers as a single object or the two sources separately may affect the results. Future observations should focus on velocity dispersion and mass estimates to confirm whether these galaxies are dynamically interacting, providing insights into their eventual clustering status. For this study, we treat them as distinct objects.

Furthermore, galaxy par0953$+$5153\_1777, previously included in the $z\sim8$ bin owing to its photometric redshift ($z_{phot}=8.14$, see \citetalias{rojas-ruiz_probing_2020}), is now confirmed to belong to the higher $z\sim9$ bin with $z_{spec}=8.440$ and M$_{UV}=-20.44\pm0.41$. This is also the only galaxy from the BoRG-\jwst\ survey which belongs to the M$_{UV}=-20$ bin, for which a number density was not previously reported in \citetalias{rojas-ruiz_probing_2020}. 

Given the shifts between redshift and magnitude bins for a number of the sources from the 2426 program, we opt to rederive the $z\sim8$ and $z\sim9$ UVLFs. We follow a similar methodology to \citetalias{rojas-ruiz_probing_2020} and perform a Markov Chain Monte Carlo (MCMC) analysis using the \texttt{emcee} Python package \citep{foreman-mackey_emcee_2013}. To calculate the UVLF values and their associated uncertainties, we follow the method described in \citet{finkelstein_evolution_2015} using the Poisson likelihood statistics:
\begin{align}
C^2(\varphi) &= -2 \ln{L(\varphi})\\
C^2(\varphi) &= 2 \sum_{i} \left( N_{{\rm model}, i} - N_{{\rm obs}, i} + N_{{\rm obs}, i} \ln \left( \frac{N_{{\rm obs}, i}}{N_{{\rm model}, i}} \right)\right) 
\end{align}

where $C^2(\varphi)$ is the goodness-of-fit statistic and $L(\varphi)$ is the likelihood that the expected number of galaxies ($N_{\rm model}$) matches the observed number ($N_{\rm obs}$).

\begin{figure*}
\centering
\includegraphics[width=\linewidth]{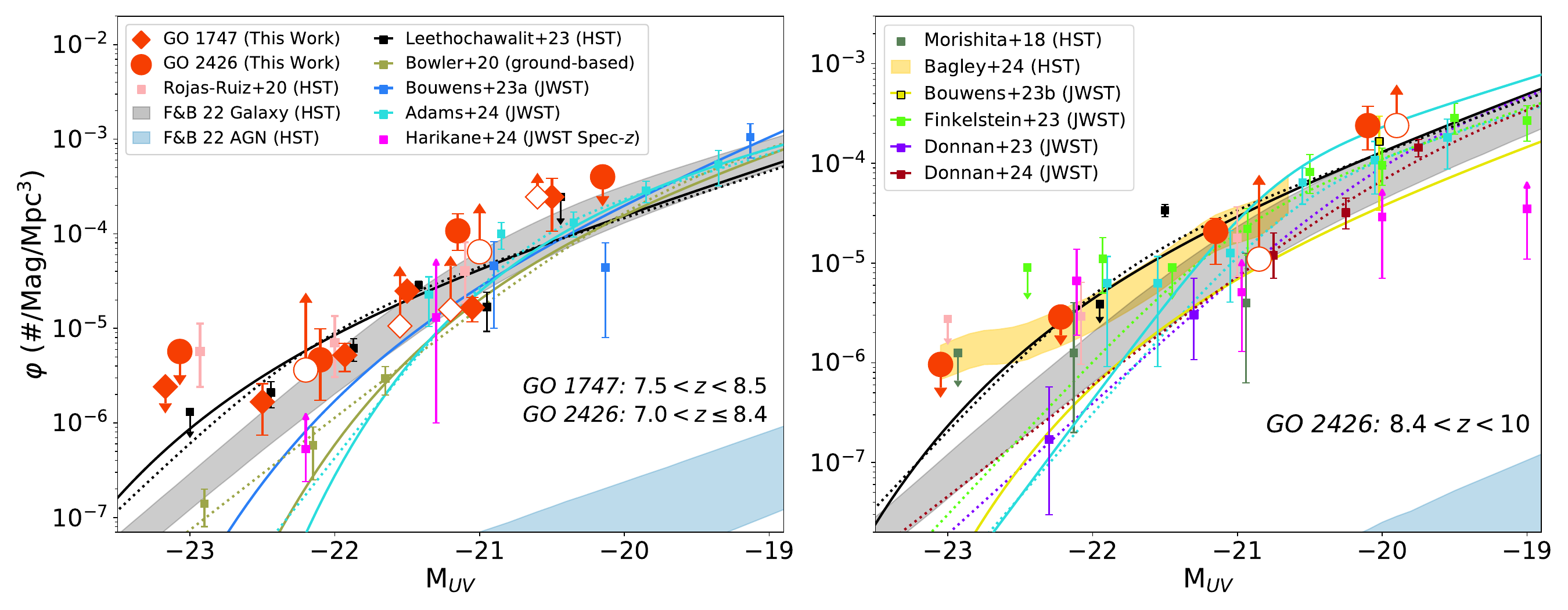}
\caption{Rest-frame UVLFs at $z\sim8$ (left) and $z\sim9$ (right) from the BoRG-\jwst\ survey program GO 1747 (orange diamonds) and GO 2426 (orange circles). We offset the number densities from the center of each M$_{\rm UV}$ bin for clarity. As a reference, we also provide the conservative hard lower limits on the number densities obtained by counting only spectroscopically confirmed galaxies, divided by the full survey volume. We include for comparison results based on previous \hst\ BoRG surveys \citep{leethochawalit_uv_2023,morishita_bright-end_2018,rojas-ruiz_probing_2020,bagley_bright_2024}, large ground-based surveys (UltraVISTA) \citep{bowler_lack_2020}, \jwst\ photometric surveys \citep{bouwens_uv_2023,bouwens_evolution_2023,donnan_evolution_2023,donnan_jwst_2024,finkelstein_complete_2024,adams_epochs_2024} and \jwst-spectroscopic results from \citet{harikane_jwst_2025} When available, the fitted Schechter UVLFs are shown as solid lines, while the double power laws are presented as dotted lines.  Our program (orange data) provides crucial bright-end constraints, especially at $z\sim8$ where previous \jwst\ results on legacy fields did not present a statistically significant number of the brightest galaxies at $M_{\rm UV} < -21.0$.}
\label{uv-lf}
\end{figure*}

In each step of the chain, we use the absolute magnitudes reported in \citetalias{roberts-borsani_borg-jwst_2025} for each galaxy confirmed from \citetalias{rojas-ruiz_probing_2020}. For each galaxy not targeted in GO 2426, we draw an absolute magnitude from a list of 1000 values randomly chosen from the cumulative distribution function of their photometric redshift probability ($P(z)$ vs. $z$) at $z>7$. This procedure allows galaxies to shift magnitude at a frequency proportional to the uncertainties in M$_\mathrm{UV}$. To account for empirical contamination, we calculate the fraction of sources from the BoRG-\jwst\ survey that are confirmed to be at lower redshift ($z\lesssim7$) relative to the total number of galaxies observed in each M$_{\rm UV}$ bin. The resulting fractions are 1, 0.6, 0.1, and 0.5 for the M$_{\rm UV}$ bins of width 1.0 centered at -23 through -20, respectively. We also rescale the contamination by the correction factor of 0.74 calculated in Section \ref{go1747}. We apply these corrections to adjust the counts of photometric objects not targeted in this survey through the MCMC. Lastly, we calculate the number densities by taking the median of the posterior distribution (from $10^5$ steps additional to $10^6$ burn-in steps) and their uncertainties from the 68\% confidence interval and report them in Table \ref{lf_table}. In Figure~\ref{uv-lf}, we show these updated number densities (filled orange circles) which largely agree within uncertainties with the photometric UVLF in \citetalias{rojas-ruiz_probing_2020} (light-pink squares), but at M$_{\rm UV}=-21$ that suffers from small number statistics having doubled in number density compared to \citetalias{rojas-ruiz_probing_2020}. We plot the number densities offset from the central absolute magnitude bin for clarity.

\section{Comparison to the Literature}\label{comp_obs}
Our UVLFs are more robust to cosmic variance owing to the nature of independent sight-line observations and, particularly at $z\sim8$, present an excess of galaxies at the bright-end compared to \jwst\ photometry as current surveys do not seem to sample the bright end well. We detail this comparison with the literature and implications of our results below. 

In Figure~\ref{uv-lf}, we show the UVLFs derived here (orange diamonds for GO 1747 and orange circles for GO 2426) and that they are largely consistent in the $z\sim8$ bin, with a minor discrepancy at M$_{\rm UV}= -21$ that could be explained by the small number statistics (one galaxy confirmed from GO 1747 compared to three galaxies from GO 2426). Although it is possible that the contamination is higher than we estimate (0.1), the few sources per magnitude bin result in significant fluctuations in the number of contaminants.
We compare our UVLFs to those previously derived from the \hst\ SuperBoRG, BoRG and WISP pure-parallel surveys \citep{morishita_bright-end_2018,rojas-ruiz_probing_2020,leethochawalit_uv_2023,bagley_bright_2024} and note that they are in good agreement at both redshift bins. 

We also compare our results to the recent \jwst/NIRCam UVLFs at $z\sim8-9$. At $z\sim8$, our number densities are consistent at the faint-end and up to M$_{UV} \gtrsim -21$ with \citet{bouwens_uv_2023}, and \citet{adams_epochs_2024}. At brighter magnitudes, we are in nice agreement with the lower limits from spectroscopically confirmed galaxies in \citet{harikane_jwst_2025}. However, for the photometrically based samples, the discrepancy is considerably higher compared to the UVLF extrapolated fits from \jwst\ results in \citet{bouwens_uv_2023}, and \citet{ adams_epochs_2024}.

Our galaxy number densities at $z\sim8$ are nearly an order of magnitude higher at (M$_{UV} \lesssim -22$) compared to the ground-based results from \citet{bowler_lack_2020}. Even our strict lower limits, based solely on spectroscopic confirmations (white symbols in Figure \ref{uv-lf}), remain significantly higher. The ground-based survey by \citet{bowler_lack_2020} covered a substantially larger area of $\sim 6$ deg$^2$ compared to the combined area of the parent programs for BoRG-\jwst\ at $z\sim8$, which is $\sim 0.4$ deg$^2$. While the larger area suggests lower cosmic variance, the BoRG-\jwst\ survey's observations from independent sight lines ensure a robust representation of galaxies at these redshifts resulting in a $\sim9\%$ lower cosmic variance effect at the brightest magnitudes (M$_{UV} \sim -23$). Possible explanations for the discrepancy may include the inherent challenges of ground-based observations in detecting significant numbers of galaxies at these bright magnitudes, or the use of more conservative selection methods to avoid contamination. This also suggests that our NIRSpec observations may capture a population of luminous galaxies not fully accounted for in previous studies.

At $z\sim9$, our measured number densities at the bright-end appear more consistent with other studies making use of \jwst\ imaging \citep{bouwens_evolution_2023,donnan_evolution_2023,donnan_jwst_2024,finkelstein_complete_2024,adams_epochs_2024}. In particular, our results are in good agreement with the spectroscopic UVLF (lower limits and data point) from \citet{harikane_jwst_2025}. Those results, derived from NIRCam imaging and spectroscopic measurements over multiple deep fields and lensing clusters (CEERS, GLASS-JWST, SMACS0723, MACS0647, GOODS-North, GOODS-South, MACS1149), imply a higher abundance of bright galaxies at the considered redshifts, more consistent with predictions from \hst\ pure-parallel surveys that benefit from reduced cosmic variance effects. This suggests that earlier studies may have underrepresented the abundance of the most luminous galaxies, aligning better with models proposing stochastic star formation processes. To fully understand the physical processes responsible for our observed number densities of these bright galaxies, in the next Section we compare to UVLFs from different analytical studies and simulations. 

Finally, while the BoRG-\textit{JWST} spectra show no evidence for strong broad-line features characteristic of type 1 AGN (see Section 4.3 in \citetalias{roberts-borsani_borg-jwst_2025}), we nonetheless compare our results to the AGN UVLF determined by \citet[][hereafter FB22]{finkelstein_coevolution_2022}, using space- and ground-based surveys predating the \jwst\ era.

The gray and blue shaded regions in Figure~\ref{uv-lf} show the galaxy and AGN luminosity functions, respectively, for both of our redshift bins. At the bright end (M$_{UV} < -21$) the number densities from our BoRG-\jwst\ work are at the upper end of the uncertainty envelope of FB22 in both redshift bins. Contamination by AGN is expected to be minimal at these bright magnitudes, based on the FB22 luminosity function. However, we caution that the FB22 AGN UVLF is constrained mostly by $z=3-6$ sources and thus the extrapolation to $z\sim8$ is forced to be smooth. The very small contribution of AGN to the total UVLF in FB22 is consistent with the lack of strong AGN features seen in our NIRSpec spectra.

\begin{figure*}
\centering
\includegraphics[width=\linewidth]{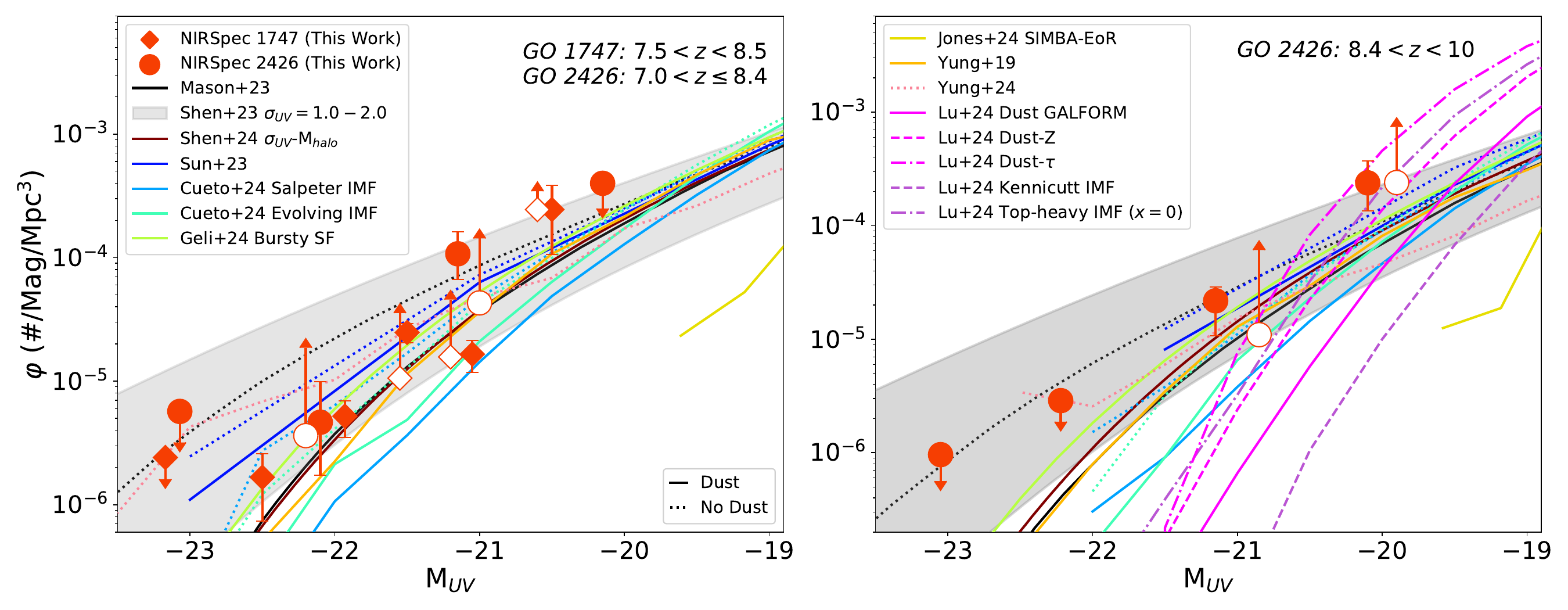}
\caption{Comparison of the UVLFs at $z\sim8$ (left) and $z\sim9$ (right) from the BoRG-\jwst\ survey to theoretical models. We include models from \citet{yung_semi-analytic_2019,yung_are_2024}, \citet{mason_brightest_2023}, \citet{shen_impact_2023,shen_early_2024}, \citet{sun_bursty_2023}, \citet{cueto_astraeus_2024},   \citet{gelli_impact_2024}, and \citet{jones_simba-eor_2024}, \citet{lu_comparison_2025}. The solid (dotted) lines show the models that (do not) include dust attenuation.}
\label{theo-lf}
\end{figure*}

\section{Comparison to Theoretical Models}\label{sims-lf}
Here we compare our derived UVLFs from BoRG-\jwst\ to the prediction of various theoretical models and simulations. Figure \ref{theo-lf} shows our observations compared to the models discussed below. 

The observed UVLFs from BoRG-\jwst\ at $z\sim8$ and $z\sim9$ agree with the analytical models by \citet{mason_brightest_2023} that consider the halo formation and variations in star formation efficiency with dust attenuation more favorably than without dust. This is similarly observed with the analytical models by \citet{yung_semi-analytic_2019} (including dust attenuation) and \citet{yung_are_2024} (no dust). \citet{shen_impact_2023} investigate the impact of UV radiation variations in star formation efficiency on the UVLF, highlighting the need for models to include stochastic star formation events. Our observations are best explained in both redshift bins with a UV scatter $\sigma_{\mathrm{UV}}= 1.0 - 2.0$ (gray shaded region). \citet{shen_early_2024} further investigates the enhanced UV variability dependent on halo mass; we show the relation for $\sigma_{\mathrm{UV}}= 0.75$ at M$_{\rm halo} \sim 10^{10.5}$ M$_\odot$ which nicely agrees with our observations, more favorably for the conservative lower limits from spectroscopically confirmed galaxies alone.
The bursty star formation models from \citet{sun_bursty_2023,gelli_impact_2024} further agree with our observations at both redshift bins. \citet{cueto_astraeus_2024} incorporate different star formation histories and initial mass functions (IMFs) following a Salpeter IMF \citep{salpeter_luminosity_1955} and an evolving IMF. For these cases, the models that do not include dust attenuation better agree with our observations.

We also compare to the SIMBA-EoR simulations using the 50 Mpc/h resolution and including dust attenuation \citep{jones_simba-eor_2024}. Although these simulations do not predict the UVLF at the bright-end (M$_{UV} <-20$) at $z\sim8-9$, the prediction at dimmer magnitudes would suggest an underestimation to our data if the trend continues at brighter magnitudes. Finally, we compare to the Durham semianalytic galaxy formation model ``GALFORM'' by \citet{lu_comparison_2025} that explore different scenarios of dust attenuation based on metallicity ($Z$) and opacity ($\tau$). They also explore a Kennicut IMF \citep{kennicutt_rate_1983} and a top-heavy IMF model. The GALFORM models present a rapid evolution of the UVLF and understimate our results at the bright end (M$_{UV} <-21$).

To further explore why our UVLFs agree better with some models with or without dust, one could examine the $\beta_{\rm UV}$ slopes. For our BoRG-\jwst\ sample of galaxies, we find that the UV continuum slopes range from $-2.5 < \beta_{\rm UV} < -1.63$ (see Table 1 in \citetalias{roberts-borsani_borg-jwst_2025}). This range of slopes is sufficiently blue, indicating low dust content. Consequently, this range of $\beta$ values does not resolve the ambiguity with the theoretical models.

Comparing our BoRG-\jwst\ UVLFs at $z\sim8$ and $z\sim9$ to theoretical models highlights the importance of measuring the bright end in order to differentiate the physical process driving galaxy evolution in the early Universe. Many of the model predictions agree with the measurements at M$_{\rm UV} \sim -20$ but fall short for brighter galaxies. Although there are multiple ways to match the bright end of the UVLF, it seems that a large scatter in UV luminosity for a given halo mass (resulting, e.g, from stochastic bursts of star formation) could do the job. As we shall see in the next Section, this scenario is qualitatively consistent with the large scatter in the observed stellar mass-to-light ratio.

\section{Stellar Mass-to-Light Ratio}\label{sec-stellar}

The M/L ratio is a powerful diagnostic of galaxy formation and evolution, and a way to connect the luminosity to the stellar mass function and the underlying dark matter halo mass function \citep[e.g.][]{behroozi_most_2018,boylan-kolchin_stress_2023, chworowsky_evidence_2024}.

Prior to \jwst, stellar masses at $z>7$ were highly uncertain owing to the limited wavelength coverage of \hst\ and the limited sensitivity of $Spitzer$. \jwst-NIRCam moved the field forward by enabling the detection of rest-frame optical colors, thus reducing the overall uncertainty on stellar masses. However, purely photometric samples still suffer from uncertainty, due to possible contamination by lower-redshift interlopers, and the effects of strong emission lines on broadband photometry.

\begin{deluxetable}{llll}
\tablecaption{Stellar Masses for BoRG-JWST}
\tablewidth{700pt}
\tablehead{
\colhead{ID} &
\colhead{$z$} &
\colhead{M$_{UV}$ }
& \colhead{log M$_*$} 
}
\startdata
1747\_1081 & 7.838 & $-$21.59$\pm$0.14 & 10.05$^{+0.03}_{-0.32}$\\
par0956$+$2847\_169 & 8.230 &  $-$22.34$\pm$0.06 & 10.02$^{+0.05}_{-0.04}$\\
par0953$+$5153\_1777 & 8.440 &  $-$20.44$\pm$0.41 & 9.68$^{+0.25}_{-0.10}$\\
par0956$+$2847\_169-neigh & 8.205 &  $-$21.45$\pm$0.09 & 9.63$^{+0.12}_{-0.04}$\\
par0953$+$5153\_1655 & 8.030 &  $-$20.68$\pm$0.19 & 9.48$^{+0.17}_{-0.24}$\\
1747\_1425 & 7.553 &  $-$21.38$\pm$0.06  & 9.44$^{+0.04}_{-0.04}$\\
par0956$+$2847\_1130 & 8.490 &  $-$20.69$\pm$0.22 & 9.37 $^{+0.13}_{-0.13}$\\
1747\_199 & 8.316 &  $-$21.28$\pm$0.17 & 9.20$^{+0.08}_{-0.03}$\\
1747\_732 & 8.226 &  $-$21.47$\pm$0.15 & 9.18 $^{+0.44}_{-0.12}$\\
par0750$+$2917\_1736 & 7.822 & $-$20.65$\pm$0.14 & 9.13$^{+0.03}_{-0.03}$\\
1747\_902 & 7.905 &  $-$21.10$\pm$0.19 & 9.02$^{+0.14}_{-0.04}$\\
1747\_817 & 7.556 &  $-$20.74$\pm$0.13 & 8.54$^{+0.14}_{-0.05}$
\enddata
\tablecomments{The redshift $z$ and M$_\mathrm{UV}$ values are from \citetalias{roberts-borsani_borg-jwst_2025}. The stellar masses log(M$_*$) that we calculate are shown in Figure \ref{ml-ratio}.}
\label{mass_table}
\end{deluxetable}

To overcome these limitations we use our NIRSpec spectra to fit for the spectral energy distribution (SED). By studying these spectral features, one can fix the redshift and break some of the degeneracies in the inferred properties such as stellar ages, metallicities, and star formation histories, which are essential for accurate stellar mass calculations (e.g., \citealt{tacchella_stellar_2022, santini_early_2023,whitler_star_2023,morishita_enhanced_2024,weibel_galaxy_2024}).

We infer the SED of individual galaxies by fitting PRISM spectra using the SED modeling code {\tt gsf} \citep[ver1.85;][]{morishita_massive_2019}. {\tt gsf} adopts nonparametric SFHs and determines an optimal combination of stellar and interstellar medium (ISM) templates. For this study, we generate templates of different ages, [10, 30, 100, 300, 1000]\,Myrs, and metallicities $\log Z_*/Z_\odot \in [-2, 0]$ at an increment of 0.1 by using {\tt fsps} \citep{conroy_propagation_2009,foreman-mackey_python-fsps_2014}. A nebular component (emission lines and continuum), which is characterized by an ionization parameter {$\log U \in [-3, -1]$} is also generated by {\tt fsps} {\citep[see also][]{byler_nebular_2017}} and added to the template after multiplication by an amplitude parameter. Dust attenuation and metallicity of the stellar templates are treated as free parameters during the fit, whereas the metallicity of the nebular component is synchronized with the metallicity of the stellar component during the fitting process. The NIRSpec PRISM spectra are scaled to the broadband fluxes (F160W, and F125W and/or F140W when available). To avoid double-counting the information, we do not include broadband fluxes in the SED fitting processes.

\begin{figure*}
\centering
\includegraphics[width=\linewidth]{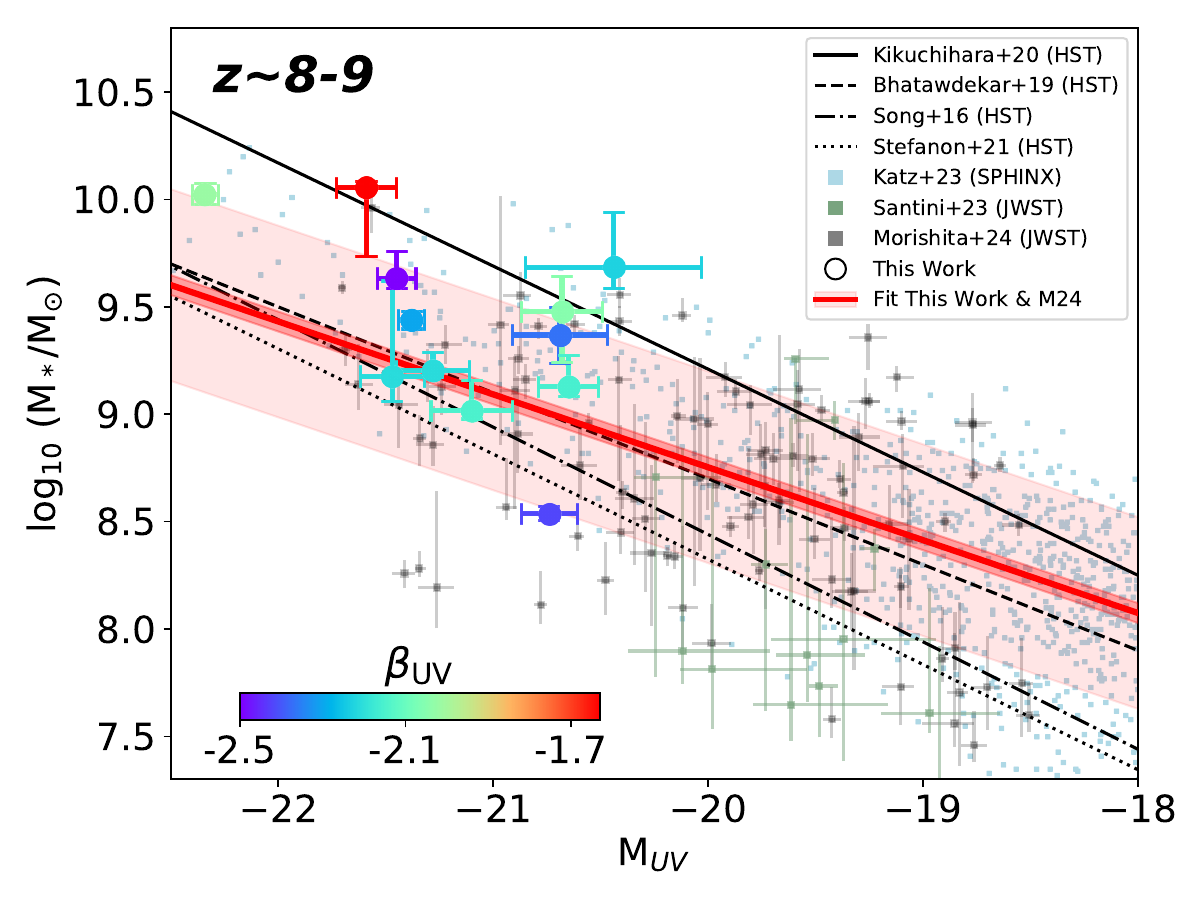}
\caption{The M/L relation for the BoRG-\jwst\ galaxies shown as circles color-coded based on the galaxy's UV continuum slope $\beta_{\rm UV}$. We fit our data jointly with stellar masses from \citet{morishita_enhanced_2024} (grey squares) that are also derived from \jwst\ observations and obtain the relation in red, where the darker-shaded region represents the 68\% interval errors and the lighter shaded area corresponds to the intrinsic scatter. We find that our relation is comparable to other M/L relations from previous \hst\ works as in \citet{song_evolution_2016,bhatawdekar_evolution_2019,kikuchihara_early_2020,stefanon_galaxy_2021}. For comparison, we also add the stellar masses from \citet{santini_early_2023} (green squares) using \jwst\ NIRCam, and \citet{katz_sphinx_2023} (blue squares) from the \text{SPHINX$^{20}$} simulations.}
\label{ml-ratio}
\end{figure*}

The calculated galaxy stellar masses ($\log M_*$) from this BoRG-\jwst\ program are presented in Table \ref{mass_table}. The relationship between $\log M_*$ and M$_{\rm UV}$ has been previously well described as a linear trend \citep[e.g.][]{song_evolution_2016,bhatawdekar_evolution_2019,kikuchihara_early_2020,stefanon_galaxy_2021}. To extend the dynamic range of our sample to fit this relation, our data is combined with the \jwst-NIRCam-based photometric sample from \citet{morishita_enhanced_2024}, which derived the stellar masses similarly through SED fitting with {\tt gsf}. We follow a Bayesian approach to linear regression as outlined in \citet{kelly_aspects_2007} using the Python package {\tt linmix}, and we show our fit in Figure \ref{ml-ratio}. 

A linear log-log relation is a good description of the data over almost three orders of magnitude in stellar mass, albeit with large scatter. The best-fit values for this relation are a slope of $-0.34\pm0.05$, with normalization (stellar mass at M$_{\mathrm{UV}} = 0$) of $1.97\pm0.05$, and an intrinsic scatter of $\sigma=0.45$. This corresponds to an almost linear relation between stellar mass and UV luminosity: M$_* \propto L_{\rm UV}^{0.85\pm0.12}$. In Figure~\ref{ml-ratio}, we also add for reference the stellar masses from \citet{santini_early_2023} using \jwst-NIRCam photometry, and recent stellar mass calculations using the \texttt{SPHINX$^{20}$} \citep{rosdahl_lyc_2022} simulations \citep{katz_sphinx_2023}. These simulations are validated against the observed photometric data on the first $\sim$ two years of \jwst.

We note that the most massive galaxies in our sample have stellar masses larger than 10$^{10}$ M$_\odot$ . This indicates that these galaxies require a high efficiency of converting gas into stars (approximately 0.2) given the abundance of dark matter halos at these redshifts and standard cosmological parameters \cite{xiao_accelerated_2024}. This high baryon star formation efficiency assumes that the IMF remains unchanged. However, if the IMF were to be more top-heavy than typically assumed, the baryon conversion efficiency could be lower \citep[e.g.][]{van_dokkum_reconciling_2024}. This potential variation in the IMF could be listed as an alternative explanation for the high efficiency observed \citep[e.g.][]{dekel_efficient_2023}. 

In Figure \ref{ml-ratio} we color-code the stellar masses (log M$_*$) based on the galaxy's UV continuum slopes ($\beta_{\rm UV}$) from Table 1 in \citetalias{roberts-borsani_borg-jwst_2025}. The most massive galaxy has the reddest $\beta$ slope, consistent with the hypothesis that it has relatively old and/or dust obscured stellar population. In contrast, galaxy par0956$+$2847\_169-neigh has the bluest slope from this sample $\beta_{\rm UV}=-2.5$ suggesting younger stellar populations or less dust and it is yet substantially massive (log M$_* =9.63^{+0.12}_{-0.04}$). Such large mass could be related to dust exchange with its neighboring galaxy and suspected merger par0956$+$2847\_169, which is among the most massive from this sample with log M$_* =10.02^{+0.05}_{-0.04}$ but redder $\beta_{\rm UV}=-2.02$ slope. Nevertheless, the $\beta_{\rm UV}$ slopes from this sample are conservative compared to bluer $\beta_{\rm UV}\sim -3$ slopes from other galaxies at similar redshifts as in \citet{morishita_enhanced_2024}. A cautionary note is that, since we use a nonparametric SFH, the masses of the low-mass blue galaxies might be underestimated if their recent star formation rate (SFR) outshines the residual light from previous star formation episodes \citep{whitler_star_2023,narayanan_outshining_2024}.

In general, our spectroscopically confirmed $z\sim8-9$ galaxies exhibit a similar linear M/L relation with the most UV luminous galaxies also having higher masses, albeit with $\sim 0.5$ dex of scatter. The large scatter and spread in $\beta_{\rm UV}$ slopes are consistent with a broad range of stellar ages and dust attenuation in the bright population. Although larger samples are needed to reach firm conclusions, this first study shows how considering the demographics with the individual properties of galaxies may help disentangle the interplay of dust, metallicity, IMF and SFHs in the early Universe.

\section{Summary}\label{discuss}

We studied the number density and M/L ratio of UV-bright galaxies at $z>7$ from BoRG-\jwst, the largest spectroscopically confirmed sample of galaxies from pure-parallel surveys. We used NIRSpec to confirm 12 galaxies at $z=7-9$ present in nine independent sight lines \citep{roberts-borsani_borg-jwst_2025}. We calculated the rest-frame UVLF at $z\sim8$ and $z\sim9$. Particularly at $z\sim8$, the abundance of bright galaxies is confirmed to be higher in parallel fields than that inferred by \hst\ studies in legacy fields. The spectroscopic nature of this study shows that the higher abundance is not due to an anomalously large number of contaminants in the parallel fields, but more likely by cosmic variance affecting the legacy fields.

Compared to recent \jwst\ studies
\citep{bouwens_evolution_2023, donnan_evolution_2023,adams_epochs_2024}, and ground-base large survey from \citet{bowler_lack_2020}, our $z\sim8$ UVLF is notably higher at the bright end (M$_{\rm UV} \lesssim -21$). We are consistent with available spectroscopic confirmations from \citet{harikane_pure_2024,harikane_jwst_2025}. At $z\sim9$, our UVLF agrees with other \jwst\ photometric findings, which survey a much smaller volume, while there is much scatter. The AGN contamination on the galaxy UVLF does not appear to be significant for either redshift bin when comparing to galaxy- and AGN- UVLF studies by \citet{finkelstein_coevolution_2022}, which is the only reference thus far at these redshifts and that is based on \hst\ data.

Our study takes us closer to pinning down the evolution of the bright end of the UVLF, which is an important diagnostic of galaxy formation and evolution, although further studies are necessary. Therefore, it is imperative to exploit the resources from \jwst\ to perform spectroscopic follow-up of UV-bright galaxies at these somewhat less explored redshifts compared to the current focus of the community at higher redshifts ($z>9$). As illustrated by our work, targeting multiple lines of sight is crucial to minimize cosmic variance.

By comparing our measurements with a range of theoretical predictions, we have illustrated the diagnostic power of the bright end of the luminosity function. Although the uncertainties are still large owing to the small number statistics, not all models match the observations. For example, models with large intrinsic scatter in UV luminosity at a given halo mass seem to do well at the bright end, consistent with the hypothesis of stochastic bursts of star formation.

We further analyzed the M/L ratio to provide insights into modeling the growth and evolution of galaxies as they turn their stellar mass into light. By combining our data of stellar masses (log M$_*$) with the \jwst\ NIRCam-based photometric sample from \citet{morishita_enhanced_2024}, we find a linear relation between  (log M$_*$) and M$_{\rm UV}$, corresponding to an almost linear relation between stellar mass and UV luminosity of the form M$_* \propto L_{\rm UV}^{0.85\pm0.12}$. The scatter around the relation is $\sim 0.5$ dex in $\log M_*$.  The large scatter is qualitatively consistent with the stochastic bursts of star formation suggested to reproduce the bright end of the galaxy UVLF.

Our result is consistent with previous studies \citep[e.g.][]{song_evolution_2016, bhatawdekar_evolution_2019, kikuchihara_early_2020, stefanon_galaxy_2021} but more robust because it is based on data reaching longer wavelengths. Our relation agrees with published \jwst-photometric data \citep{santini_early_2023} and (log M$_*$) measurements from the \text{SPHINX$^{20}$} simulations in \citet{katz_sphinx_2023}. 

To further examine this trend, we showed that the more massive galaxies with $\gtrsim 10^{10}$ M$_*$/M$_\odot$ have redder UV continuum slopes ($\beta_{\rm UV}$) indicating older, dustier stellar populations and a high efficiency of gas conversion into stars, albeit with some scatter. Low-mass blue galaxies are subject to stellar mass underestimation due to the latest SFR outshining the older stellar populations \citep{whitler_star_2023,narayanan_outshining_2024}. Changes in the assumed IMF or star formation efficiency would lead to alternative interpretations of how early galaxies achieve high masses \citep{xiao_accelerated_2024, van_dokkum_reconciling_2024, dekel_efficient_2023}. Understanding these variations via detailed studies of individual galaxies in concert with their demographics, will help us better comprehend the processes governing early galaxy evolution and the formation of massive galaxies in the early Universe. Future papers based on this BoRG-\jwst\ survey will present a detailed analysis of the individual spectra in terms of the galaxy's stellar populations, Ly$\alpha$ transmission, and intervening intergalactic medium. 
 
%====================================================
%============ Acknowledgments ==========================
%====================================================
\acknowledgements

We thank the anonymous referee for the constructive feedback. We thank E. Cueto, V. Gelli and E. Jones, X. Shen, and G. Sun for sharing their data. We are grateful to A. Calabr\'o, M. Castellano, and P. Santini, for insightful discussions and suggestions.

This work is based on observations made with the NASA/ESA/CSA James Webb Space Telescope. The data were obtained from the Mikulski Archive for Space Telescopes at the Space Telescope Science Institute, which is operated by the Association of Universities for Research in Astronomy, Inc., under NASA contract NAS 5-03127 for \jwst. The specific observations analyzed can be accessed via \dataset[doi: 10.17909/9s79-m291]{https://doi.org/10.17909/9s79-m291}. These observations are associated with programs \# 1747 and 2426. Funding from NASA through STScI awards JWST-GO-01747 and JWST-GO-02426 is gratefully acknowledged.

This research was funded in part by the Swiss National Science Foundation (SNSF) [Grant number 210558]. 
CM acknowledges support by the VILLUM FONDEN under grant 37459 and the Carlsberg Foundation under grant CF22-1322. The Cosmic Dawn Center (DAWN) is funded by the Danish National Research Foundation under grant DNRF140.  MT acknowledges  support by the Australian Research Council
Centre of Excellence for All Sky Astrophysics in 3 Dimensions
(ASTRO 3D), through project No. CE170100013. MS acknowledges support for this work under NASA grant 80NSSC22K1294.

The Flatiron Institute is a division of the Simons Foundation.

\facilities{\jwst}
\software{Astropy \citep{the_astropy_collaboration_astropy_2013},
          emcee \citep{foreman-mackey_emcee_2013},
          gsf \citep{morishita_massive_2019},
          Matplotlib \citep{hunter_matplotlib_2007},
          Numpy \citep{harris_array_2020},
          SciPy \citep{virtanen_scipy_2020}
      }
          
%=====================================================
%============ References ===========================
%=====================================================
\bibliographystyle{yahapj}
\bibliography{references.bib}      

\begin{thebibliography}{}
\providecommand\natexlab[1]{#1}
\providecommand\JournalTitle[1]{#1}

\bibitem[{Adamo {et~al.}(2024)Adamo, Atek, Bagley, Bañados, Barrow, Berg, Bezanson, Bradač, Brammer, Carnall, Chisholm, Coe, Dayal, Eisenstein, Eldridge, Ferrara, Fujimoto, de~Graaff, Habouzit, Hutchison, Kartaltepe, Kassin, Kriek, Labbé, Maiolino, Marques-Chaves, Maseda, Mason, Matthee, McQuinn, Meynet, Naidu, Oesch, Pentericci, Pérez-González, Rigby, Roberts-Borsani, Schaerer, Shapley, Stark, Stiavelli, Strom, Vanzella, Wang, Wilkins, Williams, Willott, Wylezalek, \& Nota}]{adamo_first_2024}
Adamo, A., Atek, H., Bagley, M.~B., {et~al.} 2024, \href{http://dx.doi.org/10.48550/arXiv.2405.21054}{\JournalTitle{2024arXiv240521054A}}

\bibitem[{Adams {et~al.}(2024)Adams, Conselice, Austin, Harvey, Ferreira, Trussler, Juodžbalis, Li, Windhorst, Cohen, Jansen, Summers, Tompkins, Driver, Robotham, D'Silva, Yan, Coe, Frye, Grogin, Koekemoer, Marshall, Pirzkal, Ryan, Maksym, Rutkowski, Willmer, Hammel, Nonino, Bhatawdekar, Wilkins, Bradley, Broadhurst, Cheng, Dole, Hathi, \& Zitrin}]{adams_epochs_2024}
Adams, N.~J., Conselice, C.~J., Austin, D., {et~al.} 2024, \href{http://dx.doi.org/10.3847/1538-4357/ad2a7b}{\JournalTitle{ApJ}, 965, 169}

\bibitem[{Arrabal~Haro {et~al.}(2023)Arrabal~Haro, Dickinson, Finkelstein, Kartaltepe, Donnan, Burgarella, Carnall, Cullen, Dunlop, Fernández, Fujimoto, Jung, Krips, Larson, Papovich, Pérez-González, Amorín, Bagley, Buat, Casey, Chworowsky, Cohen, Ferguson, Giavalisco, Huertas-Company, Hutchison, Kocevski, Koekemoer, Lucas, McLeod, McLure, Pirzkal, Seillé, Trump, Weiner, Wilkins, \& Zavala}]{arrabal_haro_confirmation_2023}
Arrabal~Haro, P., Dickinson, M., Finkelstein, S.~L., {et~al.} 2023, \href{http://dx.doi.org/10.1038/s41586-023-06521-7}{\JournalTitle{Natur}, 622, 707}

\bibitem[{Atek {et~al.}(2010)Atek, Malkan, McCarthy, Teplitz, Scarlata, Siana, Henry, Colbert, Ross, Bridge, Bunker, Dressler, Fosbury, Martin, \& Shim}]{atek_wfc3_2010}
Atek, H., Malkan, M., McCarthy, P., {et~al.} 2010, \href{http://dx.doi.org/10.1088/0004-637X/723/1/104}{\JournalTitle{ApJ}, 723, 104}

\bibitem[{Bagley {et~al.}(2024{\natexlab{a}})Bagley, Finkelstein, Rojas-Ruiz, Diekmann, Finkelstein, Song, Papovich, Somerville, Baronchelli, \& Dai}]{bagley_bright_2024}
Bagley, M.~B., Finkelstein, S.~L., Rojas-Ruiz, S., {et~al.} 2024{\natexlab{a}}, \href{http://dx.doi.org/10.3847/1538-4357/ad09dc}{\JournalTitle{ApJ}, 961, 209}

\bibitem[{Bagley {et~al.}(2024{\natexlab{b}})Bagley, Pirzkal, Finkelstein, Papovich, Berg, Lotz, Leung, Ferguson, Koekemoer, Dickinson, Kartaltepe, Kocevski, Somerville, Yung, Backhaus, Casey, Castellano, Chávez~Ortiz, Chworowsky, Cox, Davé, Davis, Estrada-Carpenter, Fontana, Fujimoto, Gardner, Giavalisco, Grazian, Grogin, Hathi, Hutchison, Jaskot, Jung, Kewley, Kirkpatrick, Larson, Matharu, Natarajan, Pentericci, Pérez-González, Ravindranath, Rothberg, Ryan, Shen, Simons, Snyder, Trump, \& Wilkins}]{bagley_next_2024}
Bagley, M.~B., Pirzkal, N., Finkelstein, S.~L., {et~al.} 2024{\natexlab{b}}, \href{http://dx.doi.org/10.3847/2041-8213/ad2f31}{\JournalTitle{ApJ}, 965, L6}

\bibitem[{Behroozi \& Silk(2018)}]{behroozi_most_2018}
Behroozi, P., \& Silk, J. 2018, \href{http://dx.doi.org/10.1093/mnras/sty945}{\JournalTitle{MNRAS}, 477, 5382}

\bibitem[{Bernard {et~al.}(2016)Bernard, Carrasco, Trenti, Oesch, Wu, Bradley, Schmidt, Bouwens, Calvi, Mason, Stiavelli, \& Treu}]{bernard_galaxy_2016}
Bernard, S.~R., Carrasco, D., Trenti, M., {et~al.} 2016, \href{http://dx.doi.org/10.3847/0004-637X/827/1/76}{\JournalTitle{ApJ}, 827, 76}

\bibitem[{Bhatawdekar {et~al.}(2019)Bhatawdekar, Conselice, Margalef-Bentabol, \& Duncan}]{bhatawdekar_evolution_2019}
Bhatawdekar, R., Conselice, C.~J., Margalef-Bentabol, B., \& Duncan, K. 2019, \href{http://dx.doi.org/10.1093/mnras/stz866}{\JournalTitle{MNRAS}, 486, 3805}

\bibitem[{Bouwens {et~al.}(2023{\natexlab{a}})Bouwens, Illingworth, Oesch, Stefanon, Naidu, van Leeuwen, \& Magee}]{bouwens_uv_2023}
Bouwens, R., Illingworth, G., Oesch, P., {et~al.} 2023{\natexlab{a}}, \href{http://dx.doi.org/10.1093/mnras/stad1014}{\JournalTitle{MNRAS}, 523, 1009}

\bibitem[{Bouwens {et~al.}(2019)Bouwens, Stefanon, Oesch, Illingworth, Nanayakkara, Roberts-Borsani, Labbé, \& Smit}]{bouwens_newly_2019}
Bouwens, R.~J., Stefanon, M., Oesch, P.~A., {et~al.} 2019, \href{http://dx.doi.org/10.3847/1538-4357/ab24c5}{\JournalTitle{ApJ}, 880, 25}

\bibitem[{Bouwens {et~al.}(2010)Bouwens, Illingworth, Oesch, Stiavelli, van Dokkum, Trenti, Magee, Labbé, Franx, Carollo, \& Gonzalez}]{bouwens_discovery_2010}
Bouwens, R.~J., Illingworth, G.~D., Oesch, P.~A., {et~al.} 2010, \href{http://dx.doi.org/10.1088/2041-8205/709/2/L133}{\JournalTitle{ApJL}, 709, L133}

\bibitem[{Bouwens {et~al.}(2015)Bouwens, Illingworth, Oesch, Trenti, Labbé, Bradley, Carollo, van Dokkum, Gonzalez, Holwerda, Franx, Spitler, Smit, \& Magee}]{bouwens_uv_2015}
Bouwens, R.~J., Illingworth, G.~D., Oesch, P.~A., {et~al.} 2015, \href{http://dx.doi.org/10.1088/0004-637X/803/1/34}{\JournalTitle{ApJ}, 803, 34}

\bibitem[{Bouwens {et~al.}(2016)Bouwens, Oesch, Labbé, Illingworth, Fazio, Coe, Holwerda, Smit, Stefanon, van Dokkum, Trenti, Ashby, Huang, Spitler, Straatman, Bradley, \& Magee}]{bouwens_bright_2016}
Bouwens, R.~J., Oesch, P.~A., Labbé, I., {et~al.} 2016, \href{http://dx.doi.org/10.3847/0004-637X/830/2/67}{\JournalTitle{ApJ}, 830, 67}

\bibitem[{Bouwens {et~al.}(2021)Bouwens, Oesch, Stefanon, Illingworth, Labbé, Reddy, Atek, Montes, Naidu, Nanayakkara, Nelson, \& Wilkins}]{bouwens_new_2021}
Bouwens, R.~J., Oesch, P.~A., Stefanon, M., {et~al.} 2021, \href{http://dx.doi.org/10.3847/1538-3881/abf83e}{\JournalTitle{AJ}, 162, 47}

\bibitem[{Bouwens {et~al.}(2023{\natexlab{b}})Bouwens, Stefanon, Brammer, Oesch, Herard-Demanche, Illingworth, Matthee, Naidu, van Dokkum, \& van Leeuwen}]{bouwens_evolution_2023}
Bouwens, R.~J., Stefanon, M., Brammer, G., {et~al.} 2023{\natexlab{b}}, \href{http://dx.doi.org/10.1093/mnras/stad1145}{\JournalTitle{MNRAS}, 523, 1036}

\bibitem[{Bower {et~al.}(2012)Bower, Benson, \& Crain}]{bower_what_2012}
Bower, R.~G., Benson, A.~J., \& Crain, R.~A. 2012, \href{http://dx.doi.org/10.1111/j.1365-2966.2012.20516.x}{\JournalTitle{MNRAS}, 422, 2816}

\bibitem[{Bowler {et~al.}(2017)Bowler, Dunlop, McLure, \& McLeod}]{bowler_unveiling_2017}
Bowler, R. A.~A., Dunlop, J.~S., McLure, R.~J., \& McLeod, D.~J. 2017, \href{http://dx.doi.org/10.1093/mnras/stw3296}{\JournalTitle{MNRAS}, 466, 3612}

\bibitem[{Bowler {et~al.}(2020)Bowler, Jarvis, Dunlop, McLure, McLeod, Adams, Milvang-Jensen, \& McCracken}]{bowler_lack_2020}
Bowler, R. A.~A., Jarvis, M.~J., Dunlop, J.~S., {et~al.} 2020, \href{http://dx.doi.org/10.1093/mnras/staa313}{\JournalTitle{MNRAS}, 493, 2059}

\bibitem[{Boylan-Kolchin(2023)}]{boylan-kolchin_stress_2023}
Boylan-Kolchin, M. 2023, \href{http://dx.doi.org/10.1038/s41550-023-01937-7}{\JournalTitle{Nat Astron}, 7, 731}

\bibitem[{Bradley {et~al.}(2012)Bradley, Trenti, Oesch, Stiavelli, Treu, Bouwens, Shull, Holwerda, \& Pirzkal}]{bradley_brightest_2012}
Bradley, L.~D., Trenti, M., Oesch, P.~A., {et~al.} 2012, \href{http://dx.doi.org/10.1088/0004-637X/760/2/108}{\JournalTitle{ApJ}, 760, 108}

\bibitem[{Byler {et~al.}(2017)Byler, Dalcanton, Conroy, \& Johnson}]{byler_nebular_2017}
Byler, N., Dalcanton, J.~J., Conroy, C., \& Johnson, B.~D. 2017, \href{http://dx.doi.org/10.3847/1538-4357/aa6c66}{\JournalTitle{ApJ}, 840, 44}

\bibitem[{Calvi {et~al.}(2016)Calvi, Trenti, Stiavelli, Oesch, Bradley, Schmidt, Coe, Brammer, Bernard, Bouwens, Carrasco, Carollo, Holwerda, MacKenty, Mason, Shull, \& Treu}]{calvi_bright_2016}
Calvi, V., Trenti, M., Stiavelli, M., {et~al.} 2016, \href{http://dx.doi.org/10.3847/0004-637X/817/2/120}{\JournalTitle{ApJ}, 817, 120}

\bibitem[{Carniani {et~al.}(2024)Carniani, Hainline, D’Eugenio, Eisenstein, Jakobsen, Witstok, Johnson, Chevallard, Maiolino, Helton, Willott, Robertson, Alberts, Arribas, Baker, Bhatawdekar, Boyett, Bunker, Cameron, Cargile, Charlot, Curti, Curtis-Lake, Egami, Giardino, Isaak, Ji, Jones, Kumari, Maseda, Parlanti, Pérez-González, Rawle, Rieke, Rieke, Del~Pino, Saxena, Scholtz, Smit, Sun, Tacchella, Übler, Venturi, Williams, \& Willmer}]{carniani_spectroscopic_2024}
Carniani, S., Hainline, K., D’Eugenio, F., {et~al.} 2024, \href{http://dx.doi.org/10.1038/s41586-024-07860-9}{\JournalTitle{Natur}, 633, 318}

\bibitem[{Casey {et~al.}(2023)Casey, Kartaltepe, Drakos, Franco, Harish, Paquereau, Ilbert, Rose, Cox, Nightingale, Robertson, Silverman, Koekemoer, Massey, McCracken, Rhodes, Akins, Allen, Amvrosiadis, Arango-Toro, Bagley, Bongiorno, Capak, Champagne, Chartab, Chávez~Ortiz, Chworowsky, Cooke, Cooper, Darvish, Ding, Faisst, Finkelstein, Fujimoto, Gentile, Gillman, Gould, Gozaliasl, Hayward, He, Hemmati, Hirschmann, Jahnke, Jin, Khostovan, Kokorev, Lambrides, Laigle, Larson, Leung, Liu, Liaudat, Long, Magdis, Mahler, Mainieri, Manning, Maraston, Martin, McCleary, McKinney, McPartland, Mobasher, Pattnaik, Renzini, Rich, Sanders, Sattari, Scognamiglio, Scoville, Sheth, Shuntov, Sparre, Suzuki, Talia, Toft, Trakhtenbrot, Urry, Valentino, Vanderhoof, Vardoulaki, Weaver, Whitaker, Wilkins, Yang, \& Zavala}]{casey_cosmos-web_2023}
Casey, C.~M., Kartaltepe, J.~S., Drakos, N.~E., {et~al.} 2023, \href{http://dx.doi.org/10.3847/1538-4357/acc2bc}{\JournalTitle{ApJ}, 954, 31}

\bibitem[{Casey {et~al.}(2024)Casey, Akins, Shuntov, Ilbert, Paquereau, Franco, Hayward, Finkelstein, Boylan-Kolchin, Robertson, Allen, Brinch, Cooper, Ding, Drakos, Faisst, Fujimoto, Gillman, Harish, Hirschmann, Jin, Kartaltepe, Koekemoer, Kokorev, Liu, Long, Magdis, Maraston, Martin, McCracken, McKinney, Mobasher, Rhodes, Rich, Sanders, Silverman, Toft, Vijayan, Weaver, Wilkins, Yang, \& Zavala}]{casey_cosmos-web_2024}
Casey, C.~M., Akins, H.~B., Shuntov, M., {et~al.} 2024, \href{http://dx.doi.org/10.3847/1538-4357/ad2075}{\JournalTitle{ApJ}, 965, 98}

\bibitem[{Castellano {et~al.}(2024)Castellano, Napolitano, Fontana, Roberts-Borsani, Treu, Vanzella, Zavala, Arrabal~Haro, Calabrò, Llerena, Mascia, Merlin, Paris, Pentericci, Santini, Bakx, Bergamini, Cupani, Dickinson, Filippenko, Glazebrook, Grillo, Kelly, Malkan, Mason, Morishita, Nanayakkara, Rosati, Sani, Wang, \& Yoon}]{castellano_jwst_2024}
Castellano, M., Napolitano, L., Fontana, A., {et~al.} 2024, \href{http://dx.doi.org/10.3847/1538-4357/ad5f88}{\JournalTitle{ApJ}, 972, 143}

\bibitem[{Chabrier(2003)}]{chabrier_galactic_2003}
Chabrier, G. 2003, \href{http://dx.doi.org/10.1086/376392}{\JournalTitle{PASP}, 115, 763}

\bibitem[{Chworowsky {et~al.}(2024)Chworowsky, Finkelstein, Boylan-Kolchin, McGrath, Iyer, Papovich, Dickinson, Taylor, Yung, Arrabal~Haro, Bagley, Backhaus, Bhatawdekar, Cheng, Cleri, Cole, Cooper, Costantin, Dekel, Franco, Fujimoto, Hayward, Holwerda, Huertas-Company, Hirschmann, Hutchison, Koekemoer, Larson, Li, Long, Lucas, Pirzkal, Rodighiero, Somerville, Vanderhoof, de~la Vega, Wilkins, Yang, \& Zavala}]{chworowsky_evidence_2024}
Chworowsky, K., Finkelstein, S.~L., Boylan-Kolchin, M., {et~al.} 2024, \href{http://dx.doi.org/10.3847/1538-3881/ad57c1}{\JournalTitle{AJ}, 168, 113}

\bibitem[{Coe {et~al.}(2013)Coe, Zitrin, Carrasco, Shu, Zheng, Postman, Bradley, Koekemoer, Bouwens, Broadhurst, Monna, Host, Moustakas, Ford, Moustakas, van~der Wel, Donahue, Rodney, Benítez, Jouvel, Seitz, Kelson, \& Rosati}]{coe_clash_2013}
Coe, D., Zitrin, A., Carrasco, M., {et~al.} 2013, \href{http://dx.doi.org/10.1088/0004-637X/762/1/32}{\JournalTitle{ApJ}, 762, 32}

\bibitem[{Conroy {et~al.}(2009)Conroy, Gunn, \& White}]{conroy_propagation_2009}
Conroy, C., Gunn, J.~E., \& White, M. 2009, \href{http://dx.doi.org/10.1088/0004-637X/699/1/486}{\JournalTitle{ApJ}, 699, 486}

\bibitem[{Cueto {et~al.}(2024)Cueto, Hutter, Dayal, Gottlöber, Heintz, Mason, Trebitsch, \& Yepes}]{cueto_astraeus_2024}
Cueto, E.~R., Hutter, A., Dayal, P., {et~al.} 2024, \href{http://dx.doi.org/10.1051/0004-6361/202349017}{\JournalTitle{A\&A}, 686, A138}

\bibitem[{Dalmasso {et~al.}(2024)Dalmasso, Calabrò, Leethochawalit, Vulcani, Boyett, Trenti, Treu, Castellano, Bradač, Metha, \& Santini}]{dalmasso_rate_2024}
Dalmasso, N., Calabrò, A., Leethochawalit, N., {et~al.} 2024, \href{http://dx.doi.org/10.1093/mnras/stae2064}{\JournalTitle{MNRAS}, 533, 4472}

\bibitem[{Dekel {et~al.}(2023)Dekel, Sarkar, Birnboim, Mandelker, \& Li}]{dekel_efficient_2023}
Dekel, A., Sarkar, K.~C., Birnboim, Y., Mandelker, N., \& Li, Z. 2023, \href{http://dx.doi.org/10.1093/mnras/stad1557}{\JournalTitle{MNRAS}, 523, 3201}

\bibitem[{Donnan {et~al.}(2023)Donnan, McLeod, Dunlop, McLure, Carnall, Begley, Cullen, Hamadouche, Bowler, Magee, McCracken, Milvang-Jensen, Moneti, \& Targett}]{donnan_evolution_2023}
Donnan, C.~T., McLeod, D.~J., Dunlop, J.~S., {et~al.} 2023, \href{http://dx.doi.org/10.1093/mnras/stac3472}{\JournalTitle{MNRAS}, 518, 6011}

\bibitem[{Donnan {et~al.}(2024)Donnan, McLure, Dunlop, McLeod, Magee, Arellano-Córdova, Barrufet, Begley, Bowler, Carnall, Cullen, Ellis, Fontana, Illingworth, Grogin, Hamadouche, Koekemoer, Liu, Mason, Santini, \& Stanton}]{donnan_jwst_2024}
Donnan, C.~T., McLure, R.~J., Dunlop, J.~S., {et~al.} 2024, \href{http://dx.doi.org/10.1093/mnras/stae2037}{\JournalTitle{MNRAS}, 533, 3222}

\bibitem[{Duan {et~al.}(2025)Duan, Conselice, Li, Austin, Harvey, Adams, Duncan, Trussler, Ferreira, Westcott, Harris, Windhorst, Holwerda, Broadhurst, Coe, Cohen, Du, Driver, Frye, Grogin, Hathi, Jansen, Koekemoer, Marshall, Nonino, Ortiz, Pirzkal, Robotham, Ryan, Summers, D'Silva, Willmer, \& Yan}]{duan_galaxy_2025}
Duan, Q., Conselice, C.~J., Li, Q., {et~al.} 2025, \href{http://dx.doi.org/10.1093/mnras/staf638}{\JournalTitle{MNRAS}}

\bibitem[{Eisenstein {et~al.}(2023)Eisenstein, Willott, Alberts, Arribas, Bonaventura, Bunker, Cameron, Carniani, Charlot, Curtis-Lake, D'Eugenio, Endsley, Ferruit, Giardino, Hainline, Hausen, Jakobsen, Johnson, Maiolino, Rieke, Rieke, Rix, Robertson, Stark, Tacchella, Williams, Willmer, Baker, Baum, Bhatawdekar, Boyett, Chen, Chevallard, Circosta, Curti, Danhaive, DeCoursey, de~Graaff, Dressler, Egami, Helton, Hviding, Ji, Jones, Kumari, Lützgendorf, Laseter, Looser, Lyu, Maseda, Nelson, Parlanti, Perna, Puskás, Rawle, Rodríguez Del~Pino, Sandles, Saxena, Scholtz, Sharpe, Shivaei, Silcock, Simmonds, Skarbinski, Smit, Stone, Suess, Sun, Tang, Topping, Übler, Villanueva, Wallace, Whitler, Witstok, \& Woodrum}]{eisenstein_overview_2023}
Eisenstein, D.~J., Willott, C., Alberts, S., {et~al.} 2023, \href{http://dx.doi.org/10.48550/arXiv.2306.02465}{\JournalTitle{arXiv.2306.02465}}

\bibitem[{Ellis {et~al.}(2013)Ellis, McLure, Dunlop, Robertson, Ono, Schenker, Koekemoer, Bowler, Ouchi, Rogers, Curtis-Lake, Schneider, Charlot, Stark, Furlanetto, \& Cirasuolo}]{ellis_abundance_2013}
Ellis, R.~S., McLure, R.~J., Dunlop, J.~S., {et~al.} 2013, \href{http://dx.doi.org/10.1088/2041-8205/763/1/L7}{\JournalTitle{APJL}, 763, L7}

\bibitem[{Finkelstein(2016)}]{finkelstein_observational_2016}
Finkelstein, S.~L. 2016, \href{http://dx.doi.org/10.1017/pasa.2016.26}{\JournalTitle{PASA}, 33, e037}

\bibitem[{Finkelstein \& Bagley(2022)}]{finkelstein_coevolution_2022}
Finkelstein, S.~L., \& Bagley, M.~B. 2022, \href{http://dx.doi.org/10.3847/1538-4357/ac89eb}{\JournalTitle{ApJ}, 938, 25}

\bibitem[{Finkelstein {et~al.}(2012)Finkelstein, Papovich, Salmon, Finlator, Dickinson, Ferguson, Giavalisco, Koekemoer, Reddy, Bassett, Conselice, Dunlop, Faber, Grogin, Hathi, Kocevski, Lai, Lee, McLure, Mobasher, \& Newman}]{finkelstein_candels_2012}
Finkelstein, S.~L., Papovich, C., Salmon, B., {et~al.} 2012, \href{http://dx.doi.org/10.1088/0004-637X/756/2/164}{\JournalTitle{ApJ}, 756, 164}

\bibitem[{Finkelstein {et~al.}(2015)Finkelstein, Ryan, Papovich, Dickinson, Song, Somerville, Ferguson, Salmon, Giavalisco, Koekemoer, Ashby, Behroozi, Castellano, Dunlop, Faber, Fazio, Fontana, Grogin, Hathi, Jaacks, Kocevski, Livermore, McLure, Merlin, Mobasher, Newman, Rafelski, Tilvi, \& Willner}]{finkelstein_evolution_2015}
Finkelstein, S.~L., Ryan, R.~E., Papovich, C., {et~al.} 2015, \href{http://dx.doi.org/10.1088/0004-637X/810/1/71}{\JournalTitle{ApJ}, 810, 71}

\bibitem[{Finkelstein {et~al.}(2022)Finkelstein, Bagley, Song, Larson, Papovich, Dickinson, Finkelstein, Koekemoer, Pirzkal, Somerville, Yung, Behroozi, Ferguson, Giavalisco, Grogin, Hathi, Hutchison, Jung, Kocevski, Kawinwanichakij, Rojas-Ruiz, Ryan, Snyder, \& Tacchella}]{finkelstein_census_2022}
Finkelstein, S.~L., Bagley, M., Song, M., {et~al.} 2022, \href{http://dx.doi.org/10.3847/1538-4357/ac3aed}{\JournalTitle{ApJ}, 928, 52}

\bibitem[{Finkelstein {et~al.}(2024)Finkelstein, Leung, Bagley, Dickinson, Ferguson, Papovich, Akins, Arrabal~Haro, Davé, Dekel, Kartaltepe, Kocevski, Koekemoer, Pirzkal, Somerville, Yung, Amorín, Backhaus, Behroozi, Bisigello, Bromm, Casey, Chávez~Ortiz, Cheng, Chworowsky, Cleri, Cooper, Davis, de~la Vega, Elbaz, Franco, Fontana, Fujimoto, Giavalisco, Grogin, Holwerda, Huertas-Company, Hirschmann, Iyer, Jogee, Jung, Larson, Lucas, Mobasher, Morales, Morley, Mukherjee, Pérez-González, Ravindranath, Rodighiero, Rowland, Tacchella, Taylor, Trump, \& Wilkins}]{finkelstein_complete_2024}
Finkelstein, S.~L., Leung, G. C.~K., Bagley, M.~B., {et~al.} 2024, \href{http://dx.doi.org/10.3847/2041-8213/ad4495}{\JournalTitle{ApJL}, 969, L2}

\bibitem[{Foreman-Mackey {et~al.}(2013)Foreman-Mackey, Hogg, Lang, \& Goodman}]{foreman-mackey_emcee_2013}
Foreman-Mackey, D., Hogg, D.~W., Lang, D., \& Goodman, J. 2013, \href{http://dx.doi.org/10.1086/670067}{\JournalTitle{PASA}, 125, 306}

\bibitem[{Foreman-Mackey {et~al.}(2014)Foreman-Mackey, Sick, \& Johnson}]{foreman-mackey_python-fsps_2014}
Foreman-Mackey, D., Sick, J., \& Johnson, B. \href{http://dx.doi.org/10.5281/zenodo.12157}{2014}

\bibitem[{Gehrels(1986)}]{gehrels_confidence_1986}
Gehrels, N. 1986, \href{http://dx.doi.org/10.1086/164079}{\JournalTitle{ApJ}, 303, 336}

\bibitem[{Gelli {et~al.}(2024)Gelli, Mason, \& Hayward}]{gelli_impact_2024}
Gelli, V., Mason, C., \& Hayward, C.~C. 2024, \href{http://dx.doi.org/10.3847/1538-4357/ad7b36}{\JournalTitle{ApJ}, 975, 192}

\bibitem[{Grogin {et~al.}(2011)Grogin, Kocevski, Faber, Ferguson, Koekemoer, Riess, Acquaviva, Alexander, Almaini, Ashby, Barden, Bell, Bournaud, Brown, Caputi, Casertano, Cassata, Castellano, Challis, Chary, Cheung, Cirasuolo, Conselice, Cooray, Croton, Daddi, Dahlen, Davé, Mello, Dekel, Dickinson, Dolch, Donley, Dunlop, Dutton, Elbaz, Fazio, Filippenko, Finkelstein, Fontana, Gardner, Garnavich, Gawiser, Giavalisco, Grazian, Guo, Hathi, Häussler, Hopkins, Huang, Huang, Jha, Kartaltepe, Kirshner, Koo, Lai, Lee, Li, Lotz, Lucas, Madau, McCarthy, McGrath, McIntosh, McLure, Mobasher, Moustakas, Mozena, Nandra, Newman, Niemi, Noeske, Papovich, Pentericci, Pope, Primack, Rajan, Ravindranath, Reddy, Renzini, Rix, Robaina, Rodney, Rosario, Rosati, Salimbeni, Scarlata, Siana, Simard, Smidt, Somerville, Spinrad, Straughn, Strolger, Telford, Teplitz, Trump, Wel, Villforth, Wechsler, Weiner, Wiklind, Wild, Wilson, Wuyts, Yan, \& Yun}]{grogin_candels_2011}
Grogin, N.~A., Kocevski, D.~D., Faber, S.~M., {et~al.} 2011, \href{http://dx.doi.org/10.1088/0067-0049/197/2/35}{\JournalTitle{ApJS}, 197, 35}

\bibitem[{Harikane {et~al.}(2024)Harikane, Nakajima, Ouchi, Umeda, Isobe, Ono, Xu, \& Zhang}]{harikane_pure_2024}
Harikane, Y., Nakajima, K., Ouchi, M., {et~al.} 2024, \href{http://dx.doi.org/10.3847/1538-4357/ad0b7e}{\JournalTitle{ApJ}, 960, 56}

\bibitem[{Harikane {et~al.}(2025)Harikane, Inoue, Ellis, Ouchi, Nakazato, Yoshida, Ono, Sun, Sato, Ferrami, Fujimoto, Kashikawa, McLeod, Pérez-González, Sawicki, Sugahara, Xu, Yamanaka, Carnall, Cullen, Dunlop, Egami, Grogin, Isobe, Koekemoer, Laporte, Lee, Magee, Matsuo, Matsuoka, Mawatari, Nakajima, Nakane, Tamura, Umeda, \& Yanagisawa}]{harikane_jwst_2025}
Harikane, Y., Inoue, A.~K., Ellis, R.~S., {et~al.} 2025, \href{http://dx.doi.org/10.3847/1538-4357/ad9b2c}{\JournalTitle{ApJ}, 980, 138}

\bibitem[{Harris {et~al.}(2020)Harris, Millman, van~der Walt, Gommers, Virtanen, Cournapeau, Wieser, Taylor, Berg, Smith, Kern, Picus, Hoyer, van Kerkwijk, Brett, Haldane, del Río, Wiebe, Peterson, Gérard-Marchant, Sheppard, Reddy, Weckesser, Abbasi, Gohlke, \& Oliphant}]{harris_array_2020}
Harris, C.~R., Millman, K.~J., van~der Walt, S.~J., {et~al.} 2020, \href{http://dx.doi.org/10.1038/s41586-020-2649-2}{\JournalTitle{Nature}, 585, 357}

\bibitem[{Hunter(2007)}]{hunter_matplotlib_2007}
Hunter, J.~D. 2007, \href{http://dx.doi.org/10.1109/MCSE.2007.55}{\JournalTitle{CSE}, 9, 90}, conference Name: Computing in Science Engineering

\bibitem[{Jones {et~al.}(2024)Jones, Smith, Davé, Narayanan, \& Li}]{jones_simba-eor_2024}
Jones, E., Smith, B., Davé, R., Narayanan, D., \& Li, Q. 2024, \href{http://dx.doi.org/10.1093/mnras/stae2445}{\JournalTitle{MNRAS}, 535, 1293}

\bibitem[{Katz {et~al.}(2023)Katz, Rosdahl, Kimm, Blaizot, Choustikov, Farcy, Garel, Haehnelt, Michel-Dansac, \& Ocvirk}]{katz_sphinx_2023}
Katz, H., Rosdahl, J., Kimm, T., {et~al.} 2023, \href{http://dx.doi.org/10.21105/astro.2309.03269}{\JournalTitle{OJAp}, 6, 44}

\bibitem[{Kelly(2007)}]{kelly_aspects_2007}
Kelly, B.~C. 2007, \href{http://dx.doi.org/10.1086/519947}{\JournalTitle{ApJ}, 665, 1489}

\bibitem[{Kennicutt(1983)}]{kennicutt_rate_1983}
Kennicutt, Jr., R.~C. 1983, \href{http://dx.doi.org/10.1086/161261}{\JournalTitle{ApJ}, 272, 54}

\bibitem[{Kikuchihara {et~al.}(2020)Kikuchihara, Ouchi, Ono, Mawatari, Chevallard, Harikane, Kojima, Oguri, Bruzual, \& Charlot}]{kikuchihara_early_2020}
Kikuchihara, S., Ouchi, M., Ono, Y., {et~al.} 2020, \href{http://dx.doi.org/10.3847/1538-4357/ab7dbe}{\JournalTitle{ApJ}, 893, 60}

\bibitem[{Koekemoer {et~al.}(2011)Koekemoer, Faber, Ferguson, Grogin, Kocevski, Koo, Lai, Lotz, Lucas, McGrath, Ogaz, Rajan, Riess, Rodney, Strolger, Casertano, Castellano, Dahlen, Dickinson, Dolch, Fontana, Giavalisco, Grazian, Guo, Hathi, Huang, Wel, Yan, Acquaviva, Alexander, Almaini, Ashby, Barden, Bell, Bournaud, Brown, Caputi, Cassata, Challis, Chary, Cheung, Cirasuolo, Conselice, Cooray, Croton, Daddi, Davé, Mello, Ravel, Dekel, Donley, Dunlop, Dutton, Elbaz, Fazio, Filippenko, Finkelstein, Frazer, Gardner, Garnavich, Gawiser, Gruetzbauch, Hartley, Häussler, Herrington, Hopkins, Huang, Jha, Johnson, Kartaltepe, Khostovan, Kirshner, Lani, Lee, Li, Madau, McCarthy, McIntosh, McLure, McPartland, Mobasher, Moreira, Mortlock, Moustakas, Mozena, Nandra, Newman, Nielsen, Niemi, Noeske, Papovich, Pentericci, Pope, Primack, Ravindranath, Reddy, Renzini, Rix, Robaina, Rosario, Rosati, Salimbeni, Scarlata, Siana, Simard, Smidt, Snyder, Somerville, Spinrad, Straughn, Telford, Teplitz, Trump, Vargas, Villforth,
  Wagner, Wandro, Wechsler, Weiner, Wiklind, Wild, Wilson, Wuyts, \& Yun}]{koekemoer_candels_2011}
Koekemoer, A.~M., Faber, S.~M., Ferguson, H.~C., {et~al.} 2011, \href{http://dx.doi.org/10.1088/0067-0049/197/2/36}{\JournalTitle{ApJS}, 197, 36}

\bibitem[{Larson {et~al.}(2022)Larson, Finkelstein, Hutchison, Papovich, Bagley, Dickinson, Rojas-Ruiz, Ferguson, Jung, Giavalisco, Grazian, Pentericci, \& Tacchella}]{larson_searching_2022}
Larson, R.~L., Finkelstein, S.~L., Hutchison, T.~A., {et~al.} 2022, \href{http://dx.doi.org/10.3847/1538-4357/ac5dbd}{\JournalTitle{ApJ}, 930, 104}

\bibitem[{Leethochawalit {et~al.}(2023)Leethochawalit, Roberts-Borsani, Morishita, Trenti, \& Treu}]{leethochawalit_uv_2023}
Leethochawalit, N., Roberts-Borsani, G., Morishita, T., Trenti, M., \& Treu, T. 2023, \href{http://dx.doi.org/10.1093/mnras/stad2202}{\JournalTitle{MNRAS}, 524, 5454}

\bibitem[{Leung {et~al.}(2023)Leung, Bagley, Finkelstein, Ferguson, Koekemoer, Pérez-González, Morales, Kocevski, Yang, Somerville, Wilkins, Yung, Fujimoto, Larson, Papovich, Pirzkal, Berg, Lotz, Castellano, Chávez~Ortiz, Cheng, Dickinson, Giavalisco, Hathi, Hutchison, Jung, Kartaltepe, Natarajan, \& Rothberg}]{leung_ngdeep_2023}
Leung, G. C.~K., Bagley, M.~B., Finkelstein, S.~L., {et~al.} 2023, \href{http://dx.doi.org/10.3847/2041-8213/acf365}{\JournalTitle{ApJL}, 954, L46}

\bibitem[{Livermore {et~al.}(2018)Livermore, Trenti, Bradley, Bernard, Holwerda, Mason, \& Treu}]{livermore_hst_2018}
Livermore, R.~C., Trenti, M., Bradley, L.~D., {et~al.} 2018, \href{http://dx.doi.org/10.3847/2041-8213/aacd16}{\JournalTitle{ApJL}, 861, L17}

\bibitem[{Lotz {et~al.}(2017)Lotz, Koekemoer, Coe, Grogin, Capak, Mack, Anderson, Avila, Barker, Borncamp, Brammer, Durbin, Gunning, Hilbert, Jenkner, Khandrika, Levay, Lucas, MacKenty, Ogaz, Porterfield, Reid, Robberto, Royle, Smith, Storrie-Lombardi, Sunnquist, Surace, Taylor, Williams, Bullock, Dickinson, Finkelstein, Natarajan, Richard, Robertson, Tumlinson, Zitrin, Flanagan, Sembach, Soifer, \& Mountain}]{lotz_frontier_2017}
Lotz, J.~M., Koekemoer, A., Coe, D., {et~al.} 2017, \href{http://dx.doi.org/10.3847/1538-4357/837/1/97}{\JournalTitle{ApJ}, 837, 97}

\bibitem[{Lu {et~al.}(2025)Lu, Frenk, Bose, Lacey, Cole, Baugh, \& Helly}]{lu_comparison_2025}
Lu, S., Frenk, C.~S., Bose, S., {et~al.} 2025, \href{http://dx.doi.org/10.1093/mnras/stae2646}{\JournalTitle{MNRAS}, 536, 1018}

\bibitem[{Mason {et~al.}(2023)Mason, Trenti, \& Treu}]{mason_brightest_2023}
Mason, C.~A., Trenti, M., \& Treu, T. 2023, \href{http://dx.doi.org/10.1093/mnras/stad035}{\JournalTitle{MNRAS}, 521, 497}

\bibitem[{Mason {et~al.}(2019)Mason, Fontana, Treu, Schmidt, Hoag, Abramson, Amorin, Bradač, Guaita, Jones, Henry, Malkan, Pentericci, Trenti, \& Vanzella}]{mason_inferences_2019}
Mason, C.~A., Fontana, A., Treu, T., {et~al.} 2019, \href{http://dx.doi.org/10.1093/mnras/stz632}{\JournalTitle{MNRAS}, 485, 3947}

\bibitem[{McLeod {et~al.}(2016)McLeod, McLure, \& Dunlop}]{mcleod_z_2016}
McLeod, D.~J., McLure, R.~J., \& Dunlop, J.~S. 2016, \href{http://dx.doi.org/10.1093/mnras/stw904}{\JournalTitle{MNRAS}, 459, 3812}

\bibitem[{McLeod {et~al.}(2015)McLeod, McLure, Dunlop, Robertson, Ellis, \& Targett}]{mcleod_new_2015}
McLeod, D.~J., McLure, R.~J., Dunlop, J.~S., {et~al.} 2015, \href{http://dx.doi.org/10.1093/mnras/stv780}{\JournalTitle{MNRAS}, 450, 3032}

\bibitem[{Morishita(2021)}]{morishita_superborg_2021}
Morishita, T. 2021, \href{http://dx.doi.org/10.3847/1538-4365/abce67}{\JournalTitle{ApJS}, 253, 4}

\bibitem[{Morishita {et~al.}(2018)Morishita, Trenti, Stiavelli, Bradley, Coe, Oesch, Mason, Bridge, Holwerda, Livermore, Salmon, Schmidt, Shull, \& Treu}]{morishita_bright-end_2018}
Morishita, T., Trenti, M., Stiavelli, M., {et~al.} 2018, \href{http://dx.doi.org/10.3847/1538-4357/aae68c}{\JournalTitle{ApJ}, 867, 150}

\bibitem[{Morishita {et~al.}(2019)Morishita, Abramson, Treu, Brammer, Jones, Kelly, Stiavelli, Trenti, Vulcani, \& Wang}]{morishita_massive_2019}
Morishita, T., Abramson, L.~E., Treu, T., {et~al.} 2019, \href{http://dx.doi.org/10.3847/1538-4357/ab1d53}{\JournalTitle{ApJ}, 877, 141}

\bibitem[{Morishita {et~al.}(2024)Morishita, Stiavelli, Chary, Trenti, Bergamini, Chiaberge, Leethochawalit, Roberts-Borsani, Shen, \& Treu}]{morishita_enhanced_2024}
Morishita, T., Stiavelli, M., Chary, R.-R., {et~al.} 2024, \href{http://dx.doi.org/10.3847/1538-4357/ad1404}{\JournalTitle{ApJ}, 963, 9}

\bibitem[{Morishita {et~al.}(2025)Morishita, Mason, Kreilgaard, Trenti, Treu, Vulcani, Zhang, {Abdurro'uf}, Alavi, Atek, Bahé, Bradač, Bradley, Bunker, Coe, Colbert, Gelli, Hayes, Jones, Kodama, Leethochawalit, Liu, Malkan, Mehta, Metha, Newman, Rafelski, Roberts-Borsani, Rutkowski, Scarlata, Stiavelli, Sutanto, Takahashi, Teplitz, \& Wang}]{morishita_beacon_2025}
Morishita, T., Mason, C.~A., Kreilgaard, K.~C., {et~al.} 2025, \href{http://dx.doi.org/10.3847/1538-4357/adbbdc}{\JournalTitle{ApJ}, 983, 152}

\bibitem[{Narayanan {et~al.}(2024)Narayanan, Lower, Torrey, Brammer, Cui, Davé, Iyer, Li, Lovell, Sales, Stark, Marinacci, \& Vogelsberger}]{narayanan_outshining_2024}
Narayanan, D., Lower, S., Torrey, P., {et~al.} 2024, \href{http://dx.doi.org/10.3847/1538-4357/ad0966}{\JournalTitle{ApJ}, 961, 73}

\bibitem[{Oesch {et~al.}(2018)Oesch, Bouwens, Illingworth, Labbé, \& Stefanon}]{oesch_dearth_2018}
Oesch, P.~A., Bouwens, R.~J., Illingworth, G.~D., Labbé, I., \& Stefanon, M. 2018, \href{http://dx.doi.org/10.3847/1538-4357/aab03f}{\JournalTitle{ApJ}, 855, 105}

\bibitem[{Oesch {et~al.}(2010)Oesch, Bouwens, Illingworth, Carollo, Franx, Labbé, Magee, Stiavelli, Trenti, \& Dokkum}]{oesch_z_2010}
Oesch, P.~A., Bouwens, R.~J., Illingworth, G.~D., {et~al.} 2010, \href{http://dx.doi.org/10.1088/2041-8205/709/1/L16}{\JournalTitle{APJL}, 709, L16}

\bibitem[{Oesch {et~al.}(2013)Oesch, Bouwens, Illingworth, Labbé, Franx, van Dokkum, Trenti, Stiavelli, Gonzalez, \& Magee}]{oesch_probing_2013}
Oesch, P.~A., Bouwens, R.~J., Illingworth, G.~D., {et~al.} 2013, \href{http://dx.doi.org/10.1088/0004-637X/773/1/75}{\JournalTitle{ApJ}, 773, 75}

\bibitem[{Oesch {et~al.}(2014)Oesch, Bouwens, Illingworth, Labbé, Smit, Franx, van Dokkum, Momcheva, Ashby, Fazio, Huang, Willner, Gonzalez, Magee, Trenti, Brammer, Skelton, \& Spitler}]{oesch_most_2014}
Oesch, P.~A., Bouwens, R.~J., Illingworth, G.~D., {et~al.} 2014, \href{http://dx.doi.org/10.1088/0004-637X/786/2/108}{\JournalTitle{ApJ}, 786, 108}

\bibitem[{Oesch {et~al.}(2023)Oesch, Brammer, Naidu, Bouwens, Chisholm, Illingworth, Matthee, Nelson, Qin, Reddy, Shapley, Shivaei, van Dokkum, Weibel, Whitaker, Wuyts, Covelo-Paz, Endsley, Fudamoto, Giovinazzo, Herard-Demanche, Kerutt, Kramarenko, Labbe, Leonova, Lin, Magee, Marchesini, Maseda, Mason, Matharu, Meyer, Neufeld, Prieto~Lyon, Schaerer, Sharma, Shuntov, Smit, Stefanon, Wyithe, \& Xiao}]{oesch_jwst_2023}
Oesch, P.~A., Brammer, G., Naidu, R.~P., {et~al.} 2023, \href{http://dx.doi.org/10.1093/mnras/stad2411}{\JournalTitle{MNRAS}, 525, 2864}

\bibitem[{Roberts-Borsani {et~al.}(2022)Roberts-Borsani, Morishita, Treu, Leethochawalit, \& Trenti}]{roberts-borsani_physical_2022}
Roberts-Borsani, G., Morishita, T., Treu, T., Leethochawalit, N., \& Trenti, M. 2022, \href{http://dx.doi.org/10.3847/1538-4357/ac4803}{\JournalTitle{ApJ}, 927, 236}

\bibitem[{Roberts-Borsani {et~al.}(2024)Roberts-Borsani, Treu, Shapley, Fontana, Pentericci, Castellano, Morishita, Bergamini, \& Rosati}]{roberts-borsani_between_2024}
Roberts-Borsani, G., Treu, T., Shapley, A., {et~al.} 2024, \href{http://dx.doi.org/10.3847/1538-4357/ad85d3}{\JournalTitle{ApJ}, 976, 193}

\bibitem[{Roberts-Borsani {et~al.}(2025)Roberts-Borsani, Bagley, Rojas-Ruiz, Treu, Morishita, Finkelstein, Trenti, Arrabal~Haro, Bañados, Chávez~Ortiz, Chworowsky, Hutchison, Larson, Leethochawalit, Leung, Mason, Somerville, Stiavelli, Yung, Kassin, \& Soto}]{roberts-borsani_borg-jwst_2025}
Roberts-Borsani, G., Bagley, M., Rojas-Ruiz, S., {et~al.} 2025, \href{http://dx.doi.org/10.3847/1538-4357/adba60}{\JournalTitle{ApJ}, 983, 18}

\bibitem[{Robertson(2022)}]{robertson_galaxy_2022}
Robertson, B.~E. 2022, \href{http://dx.doi.org/10.1146/annurev-astro-120221-044656}{\JournalTitle{ARA\&A}, 60, 121}

\bibitem[{Rojas-Ruiz {et~al.}(2020)Rojas-Ruiz, Finkelstein, Bagley, Stevans, Finkelstein, Larson, Mechtley, \& Diekmann}]{rojas-ruiz_probing_2020}
Rojas-Ruiz, S., Finkelstein, S.~L., Bagley, M.~B., {et~al.} 2020, \href{http://dx.doi.org/10.3847/1538-4357/ab7659}{\JournalTitle{ApJ}, 891, 146}

\bibitem[{Rosdahl {et~al.}(2022)Rosdahl, Blaizot, Katz, Kimm, Garel, Haehnelt, Keating, Martin-Alvarez, Michel-Dansac, \& Ocvirk}]{rosdahl_lyc_2022}
Rosdahl, J., Blaizot, J., Katz, H., {et~al.} 2022, \href{http://dx.doi.org/10.1093/mnras/stac1942}{\JournalTitle{MNRAS}, 515, 2386}

\bibitem[{Salpeter(1955)}]{salpeter_luminosity_1955}
Salpeter, E.~E. 1955, \href{http://dx.doi.org/10.1086/145971}{\JournalTitle{ApJ}, 121, 161}

\bibitem[{Santini {et~al.}(2023)Santini, Fontana, Castellano, Leethochawalit, Trenti, Treu, Belfiori, Birrer, Bonchi, Merlin, Mason, Morishita, Nonino, Paris, Polenta, Rosati, Yang, Boyett, Bradac, Calabrò, Dressler, Glazebrook, Marchesini, Mascia, Nanayakkara, Pentericci, Roberts-Borsani, Scarlata, Vulcani, \& Wang}]{santini_early_2023}
Santini, P., Fontana, A., Castellano, M., {et~al.} 2023, \href{http://dx.doi.org/10.3847/2041-8213/ac9586}{\JournalTitle{ApJL}, 942, L27}

\bibitem[{Schmidt {et~al.}(2014)Schmidt, Treu, Trenti, Bradley, Kelly, Oesch, Holwerda, Shull, \& Stiavelli}]{schmidt_luminosity_2014}
Schmidt, K.~B., Treu, T., Trenti, M., {et~al.} 2014, \href{http://dx.doi.org/10.1088/0004-637X/786/1/57}{\JournalTitle{ApJ}, 786, 57}

\bibitem[{Shen {et~al.}(2023)Shen, Vogelsberger, Boylan-Kolchin, Tacchella, \& Kannan}]{shen_impact_2023}
Shen, X., Vogelsberger, M., Boylan-Kolchin, M., Tacchella, S., \& Kannan, R. 2023, \href{http://dx.doi.org/10.1093/mnras/stad2508}{\JournalTitle{MNRAS}, 525, 3254}

\bibitem[{Shen {et~al.}(2024)Shen, Vogelsberger, Boylan-Kolchin, Tacchella, \& Naidu}]{shen_early_2024}
Shen, X., Vogelsberger, M., Boylan-Kolchin, M., Tacchella, S., \& Naidu, R.~P. 2024, \href{http://dx.doi.org/10.1093/mnras/stae1932}{\JournalTitle{MNRAS}, 533, 3923}

\bibitem[{Somerville {et~al.}(2008)Somerville, Hopkins, Cox, Robertson, \& Hernquist}]{somerville_semi-analytic_2008}
Somerville, R.~S., Hopkins, P.~F., Cox, T.~J., Robertson, B.~E., \& Hernquist, L. 2008, \href{http://dx.doi.org/10.1111/j.1365-2966.2008.13805.x}{\JournalTitle{MNRAS}, 391, 481}

\bibitem[{Song {et~al.}(2016)Song, Finkelstein, Ashby, Grazian, Lu, Papovich, Salmon, Somerville, Dickinson, Duncan, Faber, Fazio, Ferguson, Fontana, Guo, Hathi, Lee, Merlin, \& Willner}]{song_evolution_2016}
Song, M., Finkelstein, S.~L., Ashby, M. L.~N., {et~al.} 2016, \href{http://dx.doi.org/10.3847/0004-637X/825/1/5}{\JournalTitle{ApJ}, 825, 5}

\bibitem[{Stefanon {et~al.}(2021)Stefanon, Bouwens, Labbé, Illingworth, Gonzalez, \& Oesch}]{stefanon_galaxy_2021}
Stefanon, M., Bouwens, R.~J., Labbé, I., {et~al.} 2021, \href{http://dx.doi.org/10.3847/1538-4357/ac1bb6}{\JournalTitle{ApJ}, 922, 29}

\bibitem[{Sun {et~al.}(2023)Sun, Faucher-Giguère, Hayward, Shen, Wetzel, \& Cochrane}]{sun_bursty_2023}
Sun, G., Faucher-Giguère, C.-A., Hayward, C.~C., {et~al.} 2023, \href{http://dx.doi.org/10.3847/2041-8213/acf85a}{\JournalTitle{ApJL}, 955, L35}

\bibitem[{Tacchella {et~al.}(2022)Tacchella, Finkelstein, Bagley, Dickinson, Ferguson, Giavalisco, Graziani, Grogin, Hathi, Hutchison, Jung, Koekemoer, Larson, Papovich, Pirzkal, Rojas-Ruiz, Song, Schneider, Somerville, Wilkins, \& Yung}]{tacchella_stellar_2022}
Tacchella, S., Finkelstein, S.~L., Bagley, M., {et~al.} 2022, \href{http://dx.doi.org/10.3847/1538-4357/ac4cad}{\JournalTitle{ApJ}, 927, 170}

\bibitem[{{The Astropy Collaboration} {et~al.}(2013){The Astropy Collaboration}, Robitaille, Tollerud, Greenfield, Droettboom, Bray, Aldcroft, Davis, Ginsburg, Price-Whelan, Kerzendorf, Conley, Crighton, Barbary, Muna, Ferguson, Grollier, Parikh, Nair, Günther, Deil, Woillez, Conseil, Kramer, Turner, Singer, Fox, Weaver, Zabalza, Edwards, Azalee~Bostroem, Burke, Casey, Crawford, Dencheva, Ely, Jenness, Labrie, Lim, Pierfederici, Pontzen, Ptak, Refsdal, Servillat, \& Streicher}]{the_astropy_collaboration_astropy_2013}
{The Astropy Collaboration}, Robitaille, T.~P., Tollerud, E.~J., {et~al.} 2013, \href{http://dx.doi.org/10.1051/0004-6361/201322068}{\JournalTitle{A\&A}, 558, A33}

\bibitem[{Trapp {et~al.}(2022)Trapp, Furlanetto, \& Yang}]{trapp_framework_2022}
Trapp, A.~C., Furlanetto, S.~R., \& Yang, J. 2022, \href{http://dx.doi.org/10.1093/mnras/stab3801}{\JournalTitle{MNRAS}, 510, 4844}

\bibitem[{Trenti \& Stiavelli(2008)}]{trenti_cosmic_2008}
Trenti, M., \& Stiavelli, M. 2008, \href{http://dx.doi.org/10.1086/528674}{\JournalTitle{ApJ}, 676, 767}

\bibitem[{Trenti {et~al.}(2011)Trenti, Bradley, Stiavelli, Oesch, Treu, Bouwens, Shull, MacKenty, Carollo, \& Illingworth}]{trenti_brightest_2011}
Trenti, M., Bradley, L.~D., Stiavelli, M., {et~al.} 2011, \href{http://dx.doi.org/10.1088/2041-8205/727/2/L39}{\JournalTitle{ApJ}, 727, L39}

\bibitem[{Trenti {et~al.}(2012)Trenti, Bradley, Stiavelli, Shull, Oesch, Bouwens, Muñoz, Romano-Diaz, Treu, Shlosman, \& Carollo}]{trenti_overdensities_2012}
Trenti, M., Bradley, L.~D., Stiavelli, M., {et~al.} 2012, \href{http://dx.doi.org/10.1088/0004-637X/746/1/55}{\JournalTitle{ApJ}, 746, 55}

\bibitem[{Treu {et~al.}(2022)Treu, Roberts-Borsani, Bradac, Brammer, Fontana, Henry, Mason, Morishita, Pentericci, Wang, Acebron, Bagley, Bergamini, Belfiori, Bonchi, Boyett, Boutsia, Calabró, Caminha, Castellano, Dressler, Glazebrook, Grillo, Jacobs, Jones, Kelly, Leethochawalit, Malkan, Marchesini, Mascia, Mercurio, Merlin, Nanayakkara, Nonino, Paris, Poggianti, Rosati, Santini, Scarlata, Shipley, Strait, Trenti, Tubthong, Vanzella, Vulcani, \& Yang}]{treu_glass-jwst_2022}
Treu, T., Roberts-Borsani, G., Bradac, M., {et~al.} 2022, \href{http://dx.doi.org/10.3847/1538-4357/ac8158}{\JournalTitle{ApJ}, 935, 110}

\bibitem[{van Dokkum \& Conroy(2024)}]{van_dokkum_reconciling_2024}
van Dokkum, P., \& Conroy, C. 2024, \href{http://dx.doi.org/10.3847/2041-8213/ad77b8}{\JournalTitle{ApJ}, 973, L32}

\bibitem[{Virtanen {et~al.}(2020)Virtanen, Gommers, Oliphant, Haberland, Reddy, Cournapeau, Burovski, Peterson, Weckesser, Bright, van~der Walt, Brett, Wilson, Millman, Mayorov, Nelson, Jones, Kern, Larson, Carey, Polat, Feng, Moore, VanderPlas, Laxalde, Perktold, Cimrman, Henriksen, Quintero, Harris, Archibald, Ribeiro, Pedregosa, \& van Mulbregt}]{virtanen_scipy_2020}
Virtanen, P., Gommers, R., Oliphant, T.~E., {et~al.} 2020, \href{http://dx.doi.org/10.1038/s41592-019-0686-2}{\JournalTitle{NatMe}, 17, 261}

\bibitem[{Weibel {et~al.}(2024)Weibel, Oesch, Barrufet, Gottumukkala, Ellis, Santini, Weaver, Allen, Bouwens, Bowler, Brammer, Carnall, Cullen, Dayal, Dickinson, Donnan, Dunlop, Giavalisco, Grogin, Illingworth, Koekemoer, Labbe, Marchesini, McLeod, McLure, Naidu, Pérez-González, Shuntov, Stefanon, Toft, \& Xiao}]{weibel_galaxy_2024}
Weibel, A., Oesch, P.~A., Barrufet, L., {et~al.} 2024, \href{http://dx.doi.org/10.1093/mnras/stae1891}{\JournalTitle{MNRAS}, 533, 1808}

\bibitem[{Whitler {et~al.}(2023)Whitler, Stark, Endsley, Leja, Charlot, \& Chevallard}]{whitler_star_2023}
Whitler, L., Stark, D.~P., Endsley, R., {et~al.} 2023, \href{http://dx.doi.org/10.1093/mnras/stad004}{\JournalTitle{MNRAS}, 519, 5859}

\bibitem[{Williams {et~al.}(2025)Williams, Oesch, Weibel, Brammer, Cloonan, Whitaker, Barrufet, Bezanson, Bowler, Dayal, Franx, Greene, Hutter, Ji, Labbé, Manning, Maseda, \& Xiao}]{williams_panoramic_2025}
Williams, C.~C., Oesch, P.~A., Weibel, A., {et~al.} 2025, \href{http://dx.doi.org/10.3847/1538-4357/ad97bc}{\JournalTitle{ApJ}, 979, 140}

\bibitem[{Windhorst {et~al.}(2023)Windhorst, Cohen, Jansen, Summers, Tompkins, Conselice, Driver, Yan, Coe, Frye, Grogin, Koekemoer, Marshall, O'Brien, Pirzkal, Robotham, Ryan, Willmer, Carleton, Diego, Keel, Porto, Redshaw, Scheller, Wilkins, Willner, Zitrin, Adams, Austin, Arendt, Beacom, Bhatawdekar, Bradley, Broadhurst, Cheng, Civano, Dai, Dole, D'Silva, Duncan, Fazio, Ferrami, Ferreira, Finkelstein, Furtak, Gim, Griffiths, Hammel, Harrington, Hathi, Holwerda, Honor, Huang, Hyun, Im, Joshi, Kamieneski, Kelly, Larson, Li, Lim, Ma, Maksym, Manzoni, Meena, Milam, Nonino, Pascale, Petric, Pierel, Polletta, Röttgering, Rutkowski, Smail, Straughn, Strolger, Swirbul, Trussler, Wang, Welch, B.~Wyithe, Yun, Zackrisson, Zhang, \& Zhao}]{windhorst_jwst_2023}
Windhorst, R.~A., Cohen, S.~H., Jansen, R.~A., {et~al.} 2023, \href{http://dx.doi.org/10.3847/1538-3881/aca163}{\JournalTitle{AJ}, 165, 13}

\bibitem[{Xiao {et~al.}(2024)Xiao, Oesch, Elbaz, Bing, Nelson, Weibel, Illingworth, van Dokkum, Naidu, Daddi, Bouwens, Matthee, Wuyts, Chisholm, Brammer, Dickinson, Magnelli, Leroy, Schaerer, Herard-Demanche, Lim, Barrufet, Endsley, Fudamoto, Gómez-Guijarro, Gottumukkala, Labbé, Magee, Marchesini, Maseda, Qin, Reddy, Shapley, Shivaei, Shuntov, Stefanon, Whitaker, \& Wyithe}]{xiao_accelerated_2024}
Xiao, M., Oesch, P.~A., Elbaz, D., {et~al.} 2024, \href{http://dx.doi.org/10.1038/s41586-024-08094-5}{\JournalTitle{Nature}, 635, 311}

\bibitem[{Yan {et~al.}(2011)Yan, Yan, Zamojski, Windhorst, McCarthy, Fan, Röttgering, Koekemoer, Robertson, Davé, \& Cai}]{yan_probing_2011}
Yan, H., Yan, L., Zamojski, M.~A., {et~al.} 2011, \href{http://dx.doi.org/10.1088/2041-8205/728/1/L22}{\JournalTitle{ApJL}, 728, L22}

\bibitem[{Yung {et~al.}(2019)Yung, Somerville, Finkelstein, Popping, \& Davé}]{yung_semi-analytic_2019}
Yung, L. Y.~A., Somerville, R.~S., Finkelstein, S.~L., Popping, G., \& Davé, R. 2019, \href{http://dx.doi.org/10.1093/mnras/sty3241}{\JournalTitle{MNRAS}, 483, 2983}

\bibitem[{Yung {et~al.}(2024)Yung, Somerville, Finkelstein, Wilkins, \& Gardner}]{yung_are_2024}
Yung, L. Y.~A., Somerville, R.~S., Finkelstein, S.~L., Wilkins, S.~M., \& Gardner, J.~P. 2024, \href{http://dx.doi.org/10.1093/mnras/stad3484}{\JournalTitle{MNRAS}, 527, 5929}

\end{thebibliography}
\end{document}